\documentclass[preprint,3p,twocolumn]{elsarticle}

\usepackage{amssymb}
\usepackage{multirow,acronym}
\usepackage{natbib}
\usepackage{booktabs}
\usepackage{amsmath}
\usepackage{subfigure}
\usepackage{color}
\usepackage{xcolor}

\usepackage{hyperref}
\usepackage[T1]{fontenc}
\usepackage{graphicx}
\usepackage{bm}
\hypersetup{colorlinks=true, citecolor=blue, linkcolor=cyan, urlcolor=magenta}

\journal{Physics of the Dark Universe}

\begin{document}
\begin{frontmatter}



\title{Dark-siren Cosmology with Decihertz Gravitational-wave Detectors}


\author[DoA,KIAA,SYSU]{Muxin Liu}
\author[DoA,KIAA]{Chang Liu}
\author[SYSU]{Yi-Ming Hu\corref{cor1}}\ead{huyiming@sysu.edu.cn}
\author[KIAA,NAOC]{Lijing Shao\corref{cor1}}\ead{lshao@pku.edu.cn}
\author[DoA,KIAA]{Yacheng Kang}

\address[DoA]{Department of Astronomy, School of Physics,
Peking University, Beijing 100871, China}

\address[KIAA]{Kavli Institute for Astronomy and
Astrophysics, Peking University, Beijing 100871, China}

\address[SYSU]{School of Physics and Astronomy, Sun Yat-sen University, 
Zhuhai 519082, China}

\address[NAOC]{National Astronomical Observatories,
Chinese Academy of Sciences, Beijing 100012, China}

\cortext[cor1]{Corresponding authors}

\begin{abstract}
\Acp{GW} originated from mergers of \acp{SBBH} are considered as dark sirens in cosmology since they usually do
not have electromagnetic counterparts. In order to study cosmos with these
events, we not only need the luminosity distances extracted from GW signals, but
also require the redshift information of sources via, say,  matching GW sky
localization with galaxy catalogs.  Based on such a methodology, we explore how
well decihertz GW detectors, DO-Optimal and DECIGO, can constrain cosmological
parameters.  Using Monte-Carlo simulated dark sirens, we find that DO-Optimal
can constrain the Hubble parameter to ${\sigma_{H_0}} / {H_0}\, \lesssim 0.23\%$
when estimating $H_0$ alone, while DECIGO performs better by a factor of 5 with
${\sigma_{H_0}} / {H_0}\lesssim 0.043\%$. Such a good precision of $H_0$ will
shed light on the $H_0$ tension. For multiple-parameter estimation, 
DECIGO can still reach a  level of relative uncertainty smaller than $7\%$.
The reason why decihertz detectors perform well is explained by their large numbers of SBBH GW
events with good distance and angular resolution.
\end{abstract}




\end{frontmatter}

\acrodef{GW}{gravitational wave}
\acrodef{SBBH}{stellar-mass binary black hole}
\acrodef{BBH}{binary black hole}
\acrodef{BH}{black hole}
\acrodef{SNR}{signal-to-noise ratio}
\acrodef{FIM}{Fisher information matrix}
\acrodef{SMF}{stellar mass function}
\acrodef{EM}{electromagnetic}
\acrodef{DOs}{Decihertz Observatories}
\acrodef{DECIGO}{DECihertz laser Interferometer Gravitational wave Observatory}
\acrodef{CDF}{cumulative distribution function}
\acrodef{MDPL}{MultiDark Planck}
\acrodef{SAGE}{Semi-Analytic Galaxy Evolution}
\acrodef{TAO}{Theoretical Astrophysical Observatory}
\acrodef{SED}{Spectral Energy Distribution}
\acrodef{IMF}{Initial Mass Function}


\section{Introduction}
\label{sec1}

The first direct detection of a \ac{GW} event, namely GW150914, marked the
beginning of \ac{GW} astronomy \citep{LIGOScientific:2016aoc}. 
According to the first half of the third observing run (O3a) detected by the 
Advanced Laser Interferometer Gravitational-wave Observatory (LIGO) and Advanced 
Virgo detectors \citep{LIGOScientific:2017vwq}, there are 39 \ac{GW} candidate events
detected so far \citep[][]{LIGOScientific:2020ibl}. Although most of
them are not associated with \ac{EM} observations, with the increasing number of
detected \ac{GW} events, a statistical study of the cosmic expansion is now
possible~\citep{LIGOScientific:2021aug}. For GW events with associated EM
counterparts, because (i) GW signals can provide the luminosity distance of the
sources, and (ii) the redshift information can be obtained from the \ac{EM}
observations, one can  constrain cosmological
parameters~\citep{Schutz:1986gp,Dalal:2006qt,Holz:2005df,Nissanke:2009kt,Nissanke:2013fka}. \cite{Schutz:1986gp,Dalal:2006qt,Holz:2005df} Such \ac{GW} events are called ``standard sirens''. For
example, the first binary neutron star merger
GW170817~\citep{LIGOScientific:2017vwq} and its numerous multi-band EM
follow-ups \citep[][]{LIGOScientific:2021psn, LIGOScientific:2017zic,
LIGOScientific:2017ync, Coulter:2017wya, DES:2017kbs, Cowperthwaite:2017dyu,
Goldstein:2017mmi, Savchenko:2017ffs} became a milestone for the new era of
multi-messenger astronomy. Based on these multi-messenger observations of
GW170817, the Hubble parameter $H_0$ is constrained to be
$H_0=70^{+12}_{-8}\,{\rm km\,s^{-1}\,Mpc^{-1}}$ (68$\%$ credible interval)
\citep{LIGOScientific:2017adf}.

By contrast, we call the \ac{GW} events without \ac{EM} counterparts ``dark
sirens'' in the cosmological context. Although we lack the \ac{EM} observations
for these events, some statistic methods can still provide constraints on the
cosmological parameters. In one of the dark-siren methods, a key point is that
one can use a galaxy catalog. \citet{LIGOScientific:2018gmd} first applied
this method to GW170817 without regard to its \ac{EM} follow-ups. Their results
constrained $H_0$ to be $H_0=77^{+37}_{-18}\,{\rm km\,s^{-1}\,Mpc^{-1}}$. 
Recent studies have explored the prospects of the dark-siren cosmology using
Bayesian framework with observations
\citep[][]{LIGOScientific:2021aug,LIGOScientific:2019zcs} as well as simulated
data \citep[][]{Gray:2019ksv,Zhao:2019gyk,Zhu:2021aat,Zhu:2021bpp}. Extending
earlier methods and criteria, in this work we  discuss the constraints on the
cosmological parameters with  space-borne decihertz \ac{GW} detectors.

In order to probe the cosmological parameters, \ac{GW} signals should have high
\ac{SNR} and the inferred parameters should have good precision.
\citet{Liu:2020nwz} have shown that decihertz \ac{GW} detectors could reach high
distance resolution as well as high angular resolution, which are expected to
provide accurate parameter estimations. \citet{Isoyama:2018rjb} have also shown
that the \ac{SNR} for a GW150914-like \ac{BBH} merger event at decihertz
waveband will be greater than that in the millihertz band. All of these indicate
that space-borne decihertz \ac{GW} detectors can provide crucial scientific
value from the detection of \ac{SBBH} merger events, especially in possessing
the potential of measuring the Hubble parameter thus clarifying the Hubble
tension.\footnote{The $H_0$ tension is defined as $\sim 4\sigma$ difference of
$H_0$ between the measurement from the calibration of Cepheid variable stars
($H_0=\,73.0^{+1.4}_{-1.4}\,{\rm km\,s^{-1}\,Mpc^{-1}}$) or Type Ia supernovae
($H_0=\,73.2^{+1.3}_{-1.3}\,{\rm km\,s^{-1}\,Mpc^{-1}}$)
\citep[][]{Riess:2019cxk,Riess:2020fzl} and the extraction from the cosmic
microwave background (CMB) ($H_0=\,67.4^{+0.5}_{-0.5}\,{\rm
km\,s^{-1}\,Mpc^{-1}}$) assuming a standard cosmological model
\citep{Planck:2018vyg}.}

In this work, we attempt to explore the constraints on the cosmological
parameters with dark sirens by decihertz \ac{GW} detectors. Specifically, we
consider the \ac{DECIGO} \citep{Kawamura:2020pcg}, and 
\ac{DOs}. \ac{DOs} have two illustrative LISA\footnote{Laser Interferometer Space Antenna 
(LISA) \citep{amaro2017laser}}-like designs, the more ambitious 
DO-Optimal and the less challenging DO-Conservative \citep{Sedda:2019uro, Sedda:2021yhn}, of
which we choose DO-Optimal as a representative to study the capabilities of \ac{DOs} 
on cosmological parameter constraints. The sensitive frequency band for DO-Optimal 
ranges from 0.01 Hz to 1 Hz, while DECIGO aims to
detect \ac{GW} sources in the frequency band between 0.1 Hz and 10 Hz. Although
some estimations in the context of cosmology with \ac{DOs} have been analyzed by
\citet{Chen:2022fvf}, we consider a more realistic situation in our work,
including the redshift error caused by the peculiar velocity
\citep[][]{Gordon:2007zw,Kocsis:2005vv,He:2019dhl} and the photo-z error caused
by the photometric measurement \citep{Ilbert:2013bf}. For the luminosity
distance errors, we further consider the bias of weak gravitational lensing
\citep{Hirata:2010ba}.

The organization of this paper is as follows. We describe the methodology in
Sec.~\ref{sec2}.  In Sec.~\ref{sec8}, we illustrate the constraints on the
cosmological parameters for DO-Optimal and DECIGO.  Finally, we present our conclusions 
in Sec.~\ref{sec11}.

\section{Methodology}
\label{sec2}

Assuming a flat $\rm \Lambda CDM$ universe throughout this work, the \ac{GW}
source's luminosity distance $D_{\rm L}$ as a function of redshift $z$ can be
written as
\begin{equation}\label{eq1}
    \begin{aligned}
      {D_{\rm L}}=\frac{ c(1+z)}{ H_0} \int_0^z \frac{{\rm d}{z'}}{\sqrt{{\rm \Omega_{M}}(1+z')^3+\rm \Omega_\Lambda}}\,,
    \end{aligned}
\end{equation}
where $\rm \Omega_M$ is the present matter density fraction relative to the
critical density, $\rm \Omega_\Lambda$ is the fractional density for present
dark energy, $c$ is the speed of light, and $H_0$ is the Hubble constant.  For a
flat universe, we further have,
\begin{equation}\label{eq2}
    \begin{aligned}
      \rm \Omega_\Lambda+\Omega_{M}=1\,.
    \end{aligned}
\end{equation}
 With  known $D_{\rm L}$ and $z$ from many sources, $H_0$, $\rm \Omega_{M}$, and
 $\rm \Omega_\Lambda$ can be constrained using Eqs.~(\ref{eq1}) and (\ref{eq2}).
 As noted in the Introduction, although we can get  $D_{\rm L}$ from  \ac{GW}
 signals, the lack of \ac{EM} follow-up observations means that we miss the
 redshift information for \ac{SBBH} mergers \citep{Graham:2020gwr}. However, the
 galaxy catalog can provide the redshift information for these dark sirens in a
 statistical way. Thus in this work, based on the simulated \ac{SBBH}
 populations and related galaxy catalog, we constrain the cosmic parameters with
 the dark sirens that are to be detected by  decihertz \ac{GW} detectors.

Below we will introduce the strategy to generate \ac{SBBH} population and galaxy
catalog in Secs.~\ref{sec3} and \ref{sec4}, respectively. In Sec.~\ref{sec5} we
describe a Bayesian framework to obtain the posterior probability distributions
of  cosmological parameters. We explain the uncertainty of the redshift and the
luminosity distance in Sec.~\ref{sec6}.

\subsection{\rm Populations of \ac{SBBH}s}
\label{sec3}

\begin{figure} 
\centering    
\includegraphics[width=1.0\columnwidth]{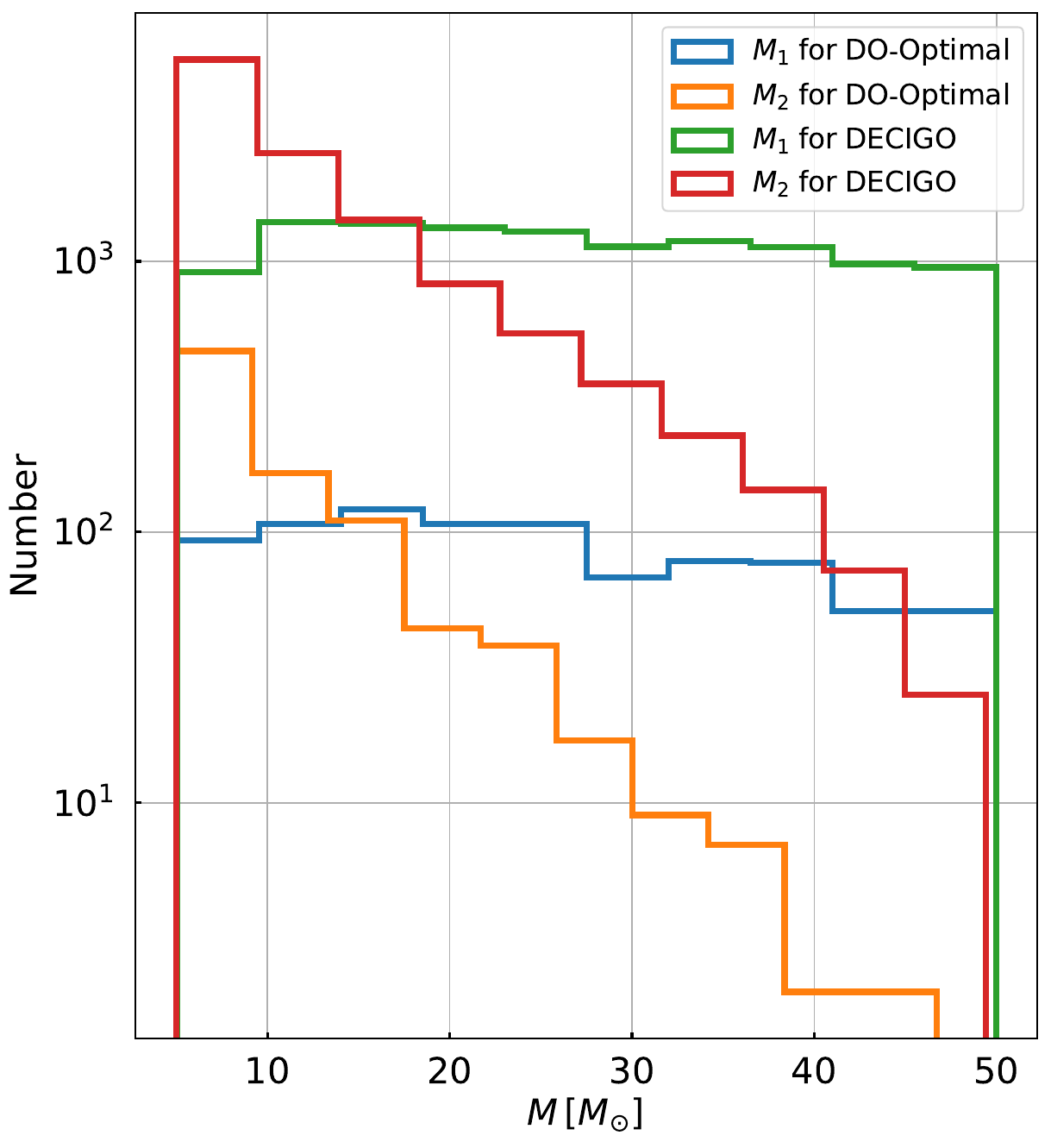}
\caption{Mass histograms for detectable \ac{GW} events in DO-Optimal and DECIGO
during their 4-year mission lifetime.}  
\label{fig:mass-distributin}     
\end{figure}
Firstly, it is important to note that we artificially set $
z_{\rm max}=1$ to be the upper detection limit in our work. This is because a larger $z$ tends 
to have a poorer localization accuracy, which will have less contribution to the cosmological 
parameter constraints. The realistic galaxy catalog is also less complete in the high-redshift 
regime. As for the mass generation of \ac{SBBH}s, we use the flat-in-log mass model to 
generate the masses for each \ac{SBBH} (see Fig.~\ref{fig:mass-distributin}) \citep{LIGOScientific:2018jsj}. 
The mass distribution for individual \ac{BH} is independently flat on the logarithmic scale as
\begin{equation}\label{eq3}
  \begin{aligned}
    {\rm P}(M_1,M_2)\propto\frac{\rm 1}{M_1M_2}\,,
  \end{aligned}
\end{equation}
where ${\rm P}(M_1,M_2)$ is the probability that the component masses of
\ac{SBBH} are $M_1$ and $M_2$, respectively ($5M_{\odot}< M_2 < M_1 < 50M_{\odot}$).
\citet{LIGOScientific:2018jsj} estimated the merger rate $\rm \mathcal{R}=
19^{+13}_{-8.2}\,Gpc^{-3}\,yr^{-1}$ under the assumption of a constant-redshift 
rate density and flat-in-log population. Conservatively, we set a 4-yr mission time 
($T_{\rm m}$) for DO-Optimal and \ac{DECIGO}. So the total number of \ac{SBBH} merger 
events is $N_{\rm tot}=\frac{4}{3}\pi (D^{\rm max}_{\rm c})^3\times\mathcal{R}\times 
T_{\rm m}$, where $D^{\rm max}_{\rm c}$ is calculated using $z_{\rm max}$ via,
\begin{equation}\label{eq5}
  \begin{aligned}
    { D_{\rm c}}=\frac{c}{H_0} \int_0^z \frac{{\rm d}z'}{\sqrt{{\rm
    \Omega_{M}}(1+z')^3+\rm \Omega_\Lambda}}\,.
  \end{aligned}
\end{equation}
And $N_{\rm tot}$ is around $1.2\times10^3$.

The sky location of \ac{GW} sources and their orbital angular momentum relative
to the Earth direction are described by $\{\overline{\theta}_{\rm
S},\overline{\phi}_{\rm S},\overline{\theta}_{\rm L},\overline{\phi}_{\rm L}\}$
in the ecliptic frame \citep{Liu:2020nwz}. We randomly generate $N_{\rm tot}$
sets of $\{\overline{\theta}_{\rm S},\overline{\phi}_{\rm
S},\overline{\theta}_{\rm L},\overline{\phi}_{\rm L}\}$ with
${\cos\,}\overline{\theta}_{\rm S},{\cos\,}\overline{\theta}_{\rm L}\in[-1,1]$
and $\overline{\phi}_{\rm S},\overline{\phi}_{\rm L}\in[0,2\pi)$. We then assign the 
\ac{GW} events to host galaxies, and replace the sky location and distance with the value 
from the nearest galaxy as its host galaxy.

In order to generate luminosity distance and redshift of \ac{SBBH}s, we assume
here that galaxies are uniformly distributed in the comoving volume. So the
probability density distribution within a spherical shell is given by
\begin{equation}\label{eq4}
    \begin{aligned}
      {\rm P}(D_{\rm c})\propto{D_{\rm c}}^2\,,
    \end{aligned}
\end{equation}
where $D_{\rm c}\le D^{\rm max}_{\rm c}$. 
The relation between luminosity distance ($D_{\rm L}$) and comoving distance
($D_{\rm c}$) is 
\begin{equation}\label{eq6}
    \begin{aligned}
      {D_{\rm L}}={(1+z)}{D_{\rm c}}\,.
    \end{aligned}
\end{equation}
We fix $H_0=67.8\,{\rm km\,s^{-1}\,Mpc^{-1}}$, $\Omega_{\rm M}=0.307$, and
$\Omega_{\Lambda}=0.693$ to be the true values in our work, 
which is derived from the $Planck$ observations \citep{Planck:2018vyg,Planck:2015fie}. 
\subsection{\rm Populations of host galaxies}
\label{sec4}

Consistent with \citet{Liu:2020nwz}, we use the 
IMRPhenomD model to produce the simulated \ac{GW} signals \citep[][]{PhysRevD.93.044007,
PhysRevD.93.044006}. The waveform is a function of the following physical parameters,
\begin{equation}\label{eq-fim-para}
  \begin{aligned}
    {\overrightarrow{\vartheta}}=\{ \mathcal{M},{D}_{\rm L},\chi_{1},\chi_{2},
    t_{\rm c},\phi_{\rm c},\eta \} \,,
  \end{aligned}
\end{equation}
where $\mathcal{M}=(M_1+M_2){\eta}^{3/5}$ is the chirp mass with the symmetric mass ratio 
$\eta=M_1M_2/{(M_1+M_2)^2}$. $t_{\rm c},\phi_{\rm c}$ are respectively the time 
and phase at coalescence and $\chi_{1},\chi_{2}$ are the dimensionless \ac{BH} spins. 
Without losing generality, we choose $t_{\rm c}=0,\phi_{\rm c}=0,\chi_{1}=0,\chi_{2}=0$ 
as fiducial values for each \ac{SBBH} \citep[][]{LIGOScientific:2020ibl}. 
And the relationship between the frequency-domain \ac{GW} signal $\widetilde{h}(f)$ 
in the detector and the incident cross and plus \ac{GW} signals is
\begin{equation}\label{eq-hf}
  \begin{aligned}
    {\widetilde{h}(f)=F^{+}(f)\widetilde{h}_{+}(f)+F^{\times}(f)\widetilde{h}_{\times}(f)}\,,
  \end{aligned}
\end{equation}
where $F^{+}(f),F^{\times}(f)$ are frequency-dependent detector pattern functions 
and $\widetilde{h}_{+}(f),\widetilde{h}_{\times}(f)$ are the source \ac{GW} waveform 
provided by the IMRPhenomD. Following \citet{Liu:2020nwz}, we use the \ac{FIM} to perform parameter
estimations for \ac{SBBH} events with \ac{SNR} larger than 10. We obtain the
angular resolution $\rm \Delta \Omega$ and the luminosity distance resolution
$\Delta D_{\rm L}$ of the \ac{SBBH} mergers. Note that the definition of 
$\rm \Delta \Omega$ is
\begin{equation}\label{eq-delta-omega}
  \begin{aligned}
    {\rm \Delta \Omega}=2\pi\sqrt{{\left(\Delta\overline{\phi}_{\rm S}
    \Delta{\rm cos}\, \overline{\theta}_{\rm S}\right)}^2
    -{\left\langle\delta\overline{\phi}_{\rm S}\delta {\rm cos}\, \overline{\theta}_{\rm S}
    \right\rangle}^2  }\,,
  \end{aligned}
\end{equation}
where $\Delta\overline{\phi}_{\rm S}$ and $\Delta{\rm cos}\, \overline{\theta}_{\rm S}$ 
are the root mean square errors of $\overline{\phi}_{\rm S}$ and ${\rm cos}\,
\overline{\theta}_{\rm S}$, respectively; $\left\langle\delta\overline{\phi}_{\rm S}\delta {\rm cos}\,
\overline{\theta}_{\rm S}\right\rangle$ is the covariance of $\overline{\phi}_{\rm S}$ 
and ${\rm cos}\, \overline{\theta}_{\rm S}$. 
More details can be seen in \citet{Liu:2020nwz}. 
Then we simulate the galaxy catalog 
with the assumption that such a galaxy catalog is completed within $z_{\rm max}=1$.  
We use flat priors for $H_0 \in [10,120] \, {\rm
km\,s^{-1}\,Mpc^{-1}}$, ${\rm \Omega_M} \in [0.1,0.5]$, and ${\rm
\Omega_\Lambda} \in [0.5,0.9]$. Combining priors and $\Delta D_{\rm L}$, we
obtain the upper (lower) limit $z^{\rm upp}$ ($z^{\rm low}$) of each \ac{SBBH}
merger. We  also obtain the upper and lower limits of the luminosity distance,
$D_{\rm L}^{\rm low}$ and $D_{\rm L}^{\rm upp}$, and the comoving distance,
$D_{\rm c}^{\rm low}$ and $D_{\rm c}^{\rm upp}$. So the comoving volumetric
error $\Delta V_{\rm c}$ of each \ac{GW} source including the host galaxy and
other candidate galaxies is given by
\begin{equation}\label{eq7}
    \begin{aligned}
      {\Delta V_{\rm c}}\approx\frac{\rm 1}{\rm 3}{\rm \Delta \Omega}\left[{
      \big( {D_{\rm c}^{\rm upp}} \big) ^3- \big( {D_{\rm c}^{\rm
      low}}\big)^3}\right]\,.
    \end{aligned}
\end{equation}
It is the volume of the frustum of a comoving cone. Given that the average
number density of the Milky-Way-like galaxy is $\sim 0.01\,\rm Mpc^{-3}$
\citep{LIGOScientific:2010nhs,Chen:2016tys}, the expected number of possible host galaxies
in each $\Delta V_{\rm c}$ can be roughly estimated as
\begin{equation}\label{eq8}
    \begin{aligned}
      {\Bar{N}_{\rm gal}}=0.01{\Delta V_{\rm c}} \,{\rm Mpc}^{-3}\,.
    \end{aligned}
\end{equation}
Note that the definition of possible host galaxies here is from the perspective
of actual detection, and it includes the host galaxy and other candidate
galaxies. We obtain the total number of galaxies $N_{\rm gal}$ from a Poisson
distribution with its mean value $\Bar{N}_{\rm gal}$. The number of other
candidate galaxies in $\Delta V_{\rm c}$ is $N_{\rm can}=N_{\rm gal}-1 $.  For
the mergers with $\Bar{N}_{\rm gal}< 1$, we can directly set $N_{\rm gal}=1$. We
generate the redshift of other candidate galaxies in the same way as \ac{SBBH}s.
In reality, galaxies are clustered on small scales rather than uniformly
distributed as assumed here \citep[][]{DES:2019ccw,DES:2020nay,Nair:2018ign}.
Clustered galaxies will provide more informative redshift distribution,
improving the constraints on cosmological parameters. Thus we are conservative
in this aspect.

As we will see later in Sec.~\ref{sec5}, we consider different weights for each
galaxy in each $\Delta V_{\rm c}$ based on their position and masses. Before
that, we first discuss how we allocate the position and mass to each galaxy. For
the position, we  get the uncertainty $\{\rm \Delta \overline{\theta}_{\rm
S},\Delta \overline{\phi}_{\rm S}\}$ by the \ac{FIM} method. We can obtain an
ellipse projected on the tangent plane of the celestial sphere, and it is
centered on the host galaxy. The principal axes of this ellipse are $\rm 3\Delta
\overline{\theta}_{\rm S}$ and $\rm 3\Delta \overline{\phi}_{\rm S}$. In this
ellipse, we randomly simulate galaxies to provide different position information
for later use.

We use a galaxy's total stellar mass as a proxy for its mass, and assume other
candidate galaxies' masses directly follow the \ac{SMF} in \citet{Kelvin_2014}.
Assuming that the formation rate of \ac{SBBH}s is uniform for all galaxies, we
expect that the more massive the galaxy is, the more likely it is the host
galaxy of the \ac{SBBH} \citep{LIGOScientific:2018gmd}. Thus, in order to highlight the contribution of the
host galaxy, we regard the mass distribution of the host galaxy as the
distribution of $\rho{M}$, where $\rho$ is the number density of galaxy and
${M}$ is the galaxy mass. The values of $\rho$ and ${M}$ are obtained from the \ac{SMF}.
Following this ``mass-weighted'' \ac{SMF}
distribution, we  generate a statistically larger mass for the host galaxy than
the candidate galaxies' masses generated from the \ac{SMF}. Note that this
treatment is crude but reasonable. At this point, we have generated the masses
of all the galaxies, and this is different from \citet{Zhu:2021aat}, and
\citet{Chen:2017gfm}.

In addition to generating host galaxy populations according to the above method, 
another realistic simulated galaxy catalog will also be considered in Sec.~\ref{secadd}. 
We will present more results in later sections.
\subsection{\rm Bayesian framework}
\label{sec5}

We use the Bayesian framework to derive the precision of the cosmological
parameters from our simulated dark sirens and related galaxy catalogs. In our
notation, a set of \ac{GW} data $\mathcal{D}_{\rm GW}\equiv\{\mathcal{D}_{\rm
GW}^1,...,\mathcal{D}_{\rm GW}^{i},...,\mathcal{D}_{\rm GW}^{N_{\rm tot}}\}$
includes $N_{\rm tot}$ \ac{GW} events, characterized by the luminosity distance
$D_{\rm L}$, the position $\{\rm \overline{\theta}_{\rm S},\overline{\phi}_{\rm
S}\}$ and so on. The posterior probability distribution for the cosmological
parameters can be estimated by
\begin{equation}\label{eq9}
    \begin{aligned}
      {\rm P}(\mathcal{H}| \mathcal{D}_{\rm GW},I)\propto {{\rm P}( \mathcal{H}|
      I)}{{\rm P}(\mathcal{D}_{\rm GW}| \mathcal{H},I)}\,,
    \end{aligned}
\end{equation}
where $\mathcal{H}\equiv\{H_0,{\rm \Omega_M,\Omega_\Lambda}\}$, ${\rm
P}(\mathcal{H}| I)$ is the prior probability distribution of $\mathcal{H}$, and
$I$ represents all the related background information. In addition, we use
${{D}_{\rm L}^i}$ to represent the true luminosity distance of the $i$-${\rm
th}$ host galaxy. We assume that it is equal to the luminosity distance of the
\ac{SBBH} merger event. The second term in the right hand side of
Eq.~(\ref{eq9}) is called the likelihood function, and it can be derived as
\begin{equation}\label{eq10}
    \begin{aligned}
      {\rm P}(\mathcal{D}_{\rm GW} | \mathcal{H},{I})=\prod_{i=1}^{N_{\rm
      tot}}{{\rm P}(\mathcal{D}_{\rm GW}^i | \mathcal{H},{I})}\,.
    \end{aligned}
\end{equation}
For each merger event, ${\rm P}(\mathcal{D}_{\rm GW}^i | \mathcal{H},{I})$ can
be expressed as
\begin{equation}\label{eq11}
    \begin{aligned}
      {\rm P}(\mathcal{D}_{\rm GW}^i | \mathcal{H},{I})=&\int {{\rm d}{D}_{\rm
      L}}\int {{\rm d}z}\int{{\rm d}M}\int{{\rm d}\Omega}\\
      &\times{{\rm P}(\mathcal{D}_{\rm GW}^i|  {{D}_{\rm
      L}},{z,{M},}\Omega,\mathcal{H},{I})} \\
      &\times{ {\rm P}({z,{M}},\Omega |  \mathcal{H},{I})}\\
      &\times{ {\rm P}({{D}_{\rm L}}| {z,{M}},\Omega,\mathcal{H},I)}\,,
    \end{aligned}
\end{equation}
where $M$ is the mass of each possible host galaxy. Adopting Gaussian noise for
\ac{GW} signals \citep{Finn:1992wt}, ${\rm P}(\mathcal{D}_{\rm GW}^i |
\mathcal{H},I)$ is assumed to follow a Gaussian distribution
$\mathcal{N}\big[\widetilde{{D}}_{\rm L}^{i},(\sigma_{{D}_{\rm L}^i,j})^2\big]$, where
$\widetilde{{D}}_{\rm L}^i$ represents the detected luminosity distance;
$\sigma_{{D}_{\rm L}^i,j}$ is the standard deviation including the error of the
redshift and the luminosity distance. We assume that $\widetilde{{D}}_{\rm L}^i$
follows $\widetilde{{D}}_{\rm L}^i \sim \mathcal{N}\big[{D}_{\rm
L}^i,{(\sigma_{D_{\rm L}}^{\rm GW})}^2\big]$.  Note that $\sigma_{D_{\rm
L}}^{\rm GW}$ here is the bias of the luminosity distance obtained from the
\ac{GW} signals alone, and it is one of the components of $\sigma_{{D}_{\rm
L}^i,j}$, to be discussed later in the next subsection. The subscript $j$
represents the $j$-th galaxy within the error volume for the $i$-th merger
event. In our notation, $j=1$ represents the host galaxy and there are $N_{\rm
can}^i$ other candidate galaxies in the $i$-th error volume. So, the redshift
information in each error volume is
$\big\{{z}_{i,1},\cdots,{z}_{i,j}, \cdots,{z}_{i,N_{\rm gal}^i}\big\}$. In addition, we
assume a $\rm \delta-$function for ${{\rm P}({{D}_{\rm L}} |
{z,M},\Omega,\mathcal{H},I)}$ via
\begin{equation}\label{eq12}
    \begin{aligned}
      {{\rm P}({{D}_{\rm L}} | {z,M},\Omega,\mathcal{H},I)}=\delta\big({D}_{\rm
      L}-{D}_{\rm L}(z,\mathcal{H})\big)\,,
    \end{aligned}
\end{equation}
where ${D}_{\rm L}(z,\mathcal{H})$ is the transformed luminosity distance of $z$
with the cosmology parameter $\mathcal{H}$ in Eq.~(\ref{eq1}).

For ${{\rm P}({z,M},\Omega |  \mathcal{H},{I})}$, we express it as
\begin{equation}\label{eq13}
    \begin{aligned}
      {\rm P}({z,M},\Omega |  \mathcal{H},{I})\propto&\sum_{j=1}^{N_{\rm
      gal}^{i}}\Big[\delta({z}-{z}_{i,j})\times{W_{\rm pos}^{i,j}(\Omega)}\\
      &\times{W_{\rm mass}^{i,j}(M)}\Big]\,,
    \end{aligned}
\end{equation}
where $\delta({z}-{z}_{i,j})$ means that the weight from the redshift of each
galaxy is the same. $W_{\rm pos}^{i,j}(\Omega)$ and $W_{\rm mass}^{i,j}(M)$
represent the weights from the position and the mass of each galaxy,
respectively. We could get the covariance matrix $\Xi_{\overline{\theta}_{\rm
S},\overline{\phi}_{\rm S}}^i$ of sky location from each merger event by the
\ac{FIM} method. The positional weight  is obtained as 
\begin{equation}\label{eq14}
    \begin{aligned}
      {f}_{\rm pos}(\overline{\theta}_{\rm S}^{i,j},\overline{\phi}_{\rm
      S}^{i,j})\propto\, & {\rm exp} \bigg[-\frac{1}{2}
      \big(\overline{\theta}_{\rm S}^{i,j}-\overline{\theta}_{\rm
      S}^{i,1},\overline{\phi}_{\rm S}^{i,j}-\overline{\phi}_{\rm S}^{i,1}
      \big)\\
      &{\Xi_{\overline{\theta}_{\rm S},\overline{\phi}_{\rm S}}^i
      \big(\overline{\theta}_{\rm S}^{i,j}-\overline{\theta}_{\rm
      S}^{i,1},\overline{\phi}_{\rm S}^{i,j}-\overline{\phi}_{\rm
      S}^{i,1}\big)^{\rm T}\bigg]}\,,
    \end{aligned}
\end{equation}
\begin{equation}\label{eq14-1}
    \begin{aligned}
      {W_{\rm pos}^{i,j}(\Omega)}\propto & {f}_{\rm pos}(\overline{\theta}_{\rm
      S},\overline{\phi}_{\rm S})\times\delta(\overline{\theta}_{\rm
      S}-\overline{\theta}_{\rm S}^{i,j})\\
      &\times\delta(\overline{\phi}_{\rm S}-\overline{\phi}_{\rm S}^{i,j})\,,
    \end{aligned}
\end{equation}
where $\{\overline{\theta}_{\rm S}^{i,j},\overline{\phi}_{\rm S}^{i,j}\}$
represent the sky location of each galaxy in $\Delta V_{\rm c}$; ${f}_{\rm
pos}(\overline{\theta}_{\rm S}^{i,j},\overline{\phi}_{\rm S}^{i,j})$ represents
the probability density at the sky location. As mentioned in Sec.~\ref{sec4},
$W_{\rm mass}^{i,j}(M)$ is defined as
\begin{equation}\label{eq15}
    \begin{aligned}
      &{W_{\rm mass}^{i,j}(M)}\propto M\times\delta({M}-{M}_{i,j})\,.
    \end{aligned}
\end{equation}
From Eqs.~(\ref{eq10}--\ref{eq15}), we can derive the likelihood for all the \ac{GW}
events as
\begin{equation}\label{eq16}
    \begin{aligned}
      {\rm P}&(\mathcal{D}_{\rm GW}|  \mathcal{H},{\rm I})\propto\prod_{i=1}^{
      N_{\rm tot}} \bigg\{ \sum_{j=1}^{N_{\rm gal}^{ i}}
      \Big[{M}_{i,j}\times{f}_{\rm pos}(\overline{\theta}_{\rm
      S}^{i,j},\overline{\phi}_{\rm S}^{i,j})\\
      &\times{\frac{1}{\sqrt{2\pi}{{\sigma_{{D}_{\rm L}^i,j}}}}{\rm exp}
      \Big(-\frac{{ \big({D}_{\rm L}({
      z}_{i,j},\mathcal{H})-\widetilde{{D}}_{\rm L}^i \big)}^2}{2
      ({\sigma_{{D}_{\rm L}^i,j}})^2} \Big)} \Big] \bigg\} \,.
    \end{aligned}
\end{equation}
According to Eqs.~(\ref{eq9}--\ref{eq16}), we calculate the posterior
probability distributions of the cosmological parameters $\mathcal{H}$. 
Note that the cosmological parameter error might originate from the EM selection 
effect. The \ac{GW} selection effect is not considered in this work because the constraints 
mainly depend on the nearest GW events, which have large SNRs and well localization or 
even the best localization. The bias introduced by the \ac{GW} selection effect is negligible 
for these events. Taking DO-Optimal as an example, from Table \ref{tab:Numberevent}, only about $10\%$ 
of the \ac{GW} events are well-localized, and about $80\%$ of these well-localized events have 
large SNRs (SNR>50). This means that the majority of the sources used for the \ac{GW} cosmology study 
are high SNR events, therefore will not suffer from \ac{GW} selection bias. Besides, 
Fig.~\ref{fig:DO-NUM} shows that DO-Optimal can detect well-localized events with $z \lesssim
0.5$. Therefore, we argue that the \ac{GW} selection effect will not introduce GW bias too much 
in the constraints of the cosmological parameters. For the influence of the EM selection effect, 
more details will be discussed in Sec.~\ref{secadd} with the simulated galaxy catalog.

We use the method of Markov Chain Monte Carlo (MCMC) to obtain the parameter estimations 
and perform a series of MCMC studies with \texttt{emcee}, a Python package that implements 
an affine-invariant MCMC ensemble sampler 
\citep[][]{Foreman_Mackey_2013,Foreman_Mackey_2019}.

\subsection{\rm Uncertainty of redshift and luminosity distance}
\label{sec6}

For the luminosity distance, we consider the following two possible errors.  (i)
We use a fitting function in \citet{Hirata:2010ba} to estimate the error from
the weak lensing effects, which is given by
\begin{equation}\label{eq19}
    \begin{aligned}
      {\sigma_{D_{\rm L}}^{\rm lens}(z)}={{\rm
      C}\;\bigg[\frac{1-(1+z)^{-\beta}}{\beta} \bigg]^{\alpha}\;D_{\rm L}}\,,
    \end{aligned}
\end{equation}
where $\rm C=0.066$, $\beta=0.25$, and $\alpha=1.8$.  (ii) As mentioned in
Sec.~\ref{sec4}, we can get $\Delta D_{\rm L}$ through \ac{FIM} for each \ac{GW}
event. So the error from the \ac{GW} measurements could be expressed as
\begin{equation}\label{eq20}
    \begin{aligned}
      {\sigma_{D_{\rm L}}^{\rm GW}(z)}={\Delta D_{\rm L}}\,.
    \end{aligned}
\end{equation}

To adequately describe the uncertainty of the redshift, we consider two kinds of
errors and then transform both uncertainties into the errors on the luminosity
distance.  (i) According to \citet{Ilbert:2013bf}, the uncertainty from the
photometric redshift is given by
\begin{equation}\label{eq17-0}
    \begin{aligned}
      {\sigma_{z}^{\rm photo}(z)}={0.008(1+z)}\,.
    \end{aligned}
\end{equation}
Then based on \citet{Kocsis:2005vv} and the chain rule, we derive
\begin{equation}\label{eq17}
    \begin{aligned}
      {\sigma_{D_{\rm L}}^{\rm photo}(z)}&=\frac {\partial {D_{\rm
      L}(z)}}{\partial {z}} {\sigma_{z}^{\rm photo}(z)} \,.
    \end{aligned}
\end{equation}
(ii) \citet[][]{Gordon:2007zw}, \citet{Kocsis:2005vv}, and \citet{He:2019dhl}
estimated the redshift error from the peculiar velocity of galaxies by
\begin{equation}\label{eq18}
    \begin{aligned}
      {\sigma_{D_{\rm L}}^{\rm pv}(z)}={ \Big[1-\frac{c(1+z)^2}{H(z)D_{\rm L}}
      \Big]\;\frac{\sqrt{\langle v^2 \rangle}}{c}\;D_{\rm L}}\,,\\
    \end{aligned}
\end{equation}
\begin{equation}\label{eq18-0}
    \begin{aligned}
      {H(z)}=H_0{\sqrt{ \Omega_{\rm M}(1+z)^3+{\rm \Omega_\Lambda}}}\,,\\
    \end{aligned}
\end{equation}
where $\sqrt{\langle v^2 \rangle}=500 \; \rm km/s$ is the root mean square
peculiar velocity of the galaxy with respect to the Hubble flow
\citep[][]{Zhu:2021aat,He:2019dhl}.

Combining all the errors mentioned above, the total uncertainty of the
luminosity distance of the $j$-th possible host galaxies corresponding to the 
$i$-th \ac{GW} event is expressed as
\begin{equation}\label{eq21}
    \begin{aligned}
      &{\sigma_{{D}_{\rm L}^i,j}({z}_{i,j})}={\sqrt{{(\sigma_{D_{\rm L}}^{\rm
      photo})}^2+{(\sigma_{D_{\rm L}}^{\rm pv})}^2+{(\sigma_{D_{\rm L}}^{\rm
      lens})}^2+{(\sigma_{D_{\rm L}}^{\rm GW})}^2}}\,.
    \end{aligned}
\end{equation}

\section{Results}
\label{sec8}

In this section, we compare the constraints on the cosmology parameters with
dark sirens between DO-Optimal and DECIGO. Sec.~\ref{sec9} gives results for
a single parameter $H_0$, and Sec.~\ref{sec10} gives results for multiple
parameters. Sec.~\ref{secadd} gives results for parameters by using a realistic
simulated galaxy catalog.

\begin{figure*}
  \centering
  \subfigure[DO-Optimal]{
      \label{fig:DO-NUM}     
      \includegraphics[width=1\columnwidth]{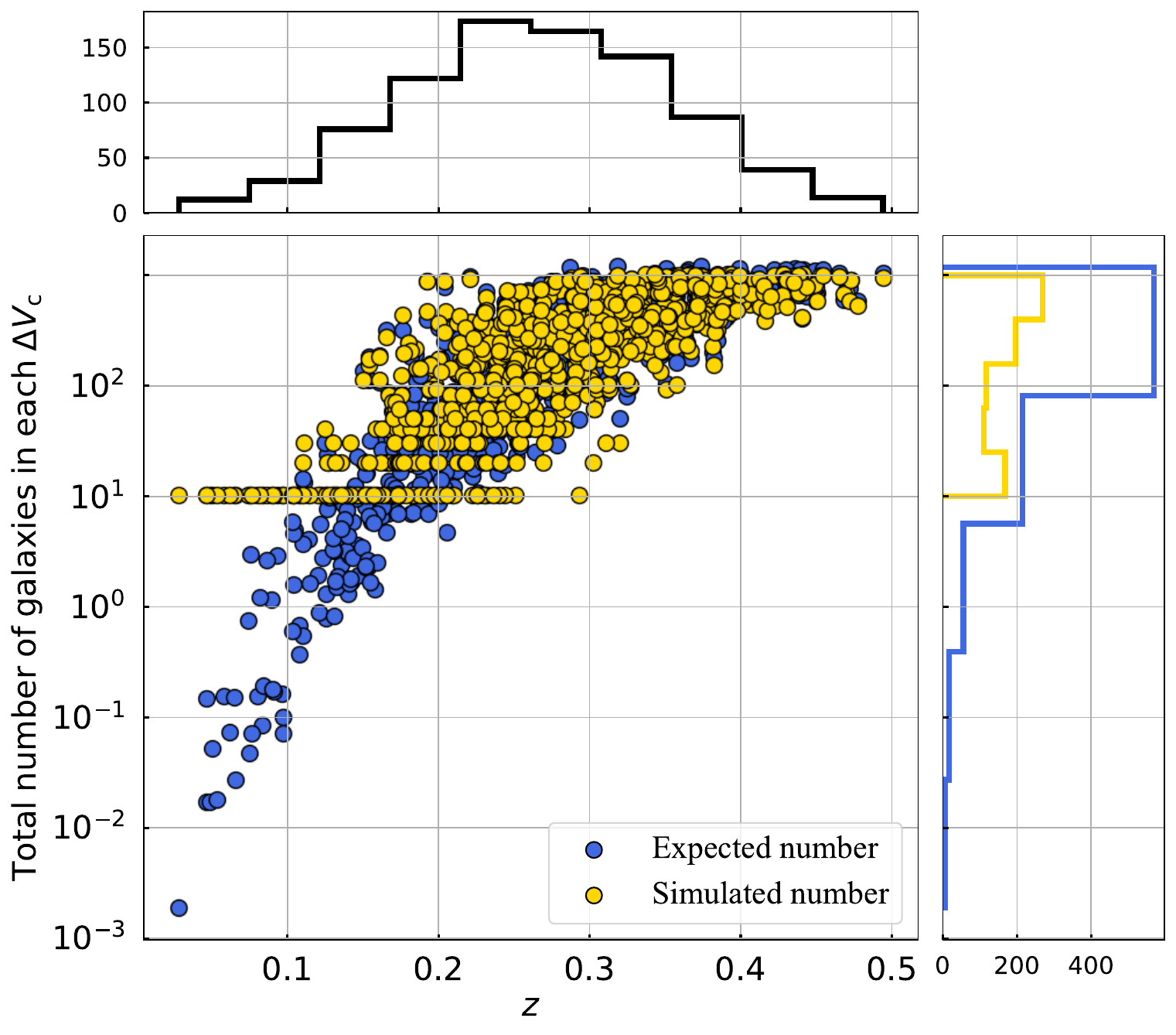}
  }
  \subfigure[DECIGO]{ 
      \label{fig:DECIGO-NUM}
      \includegraphics[width=1\columnwidth]{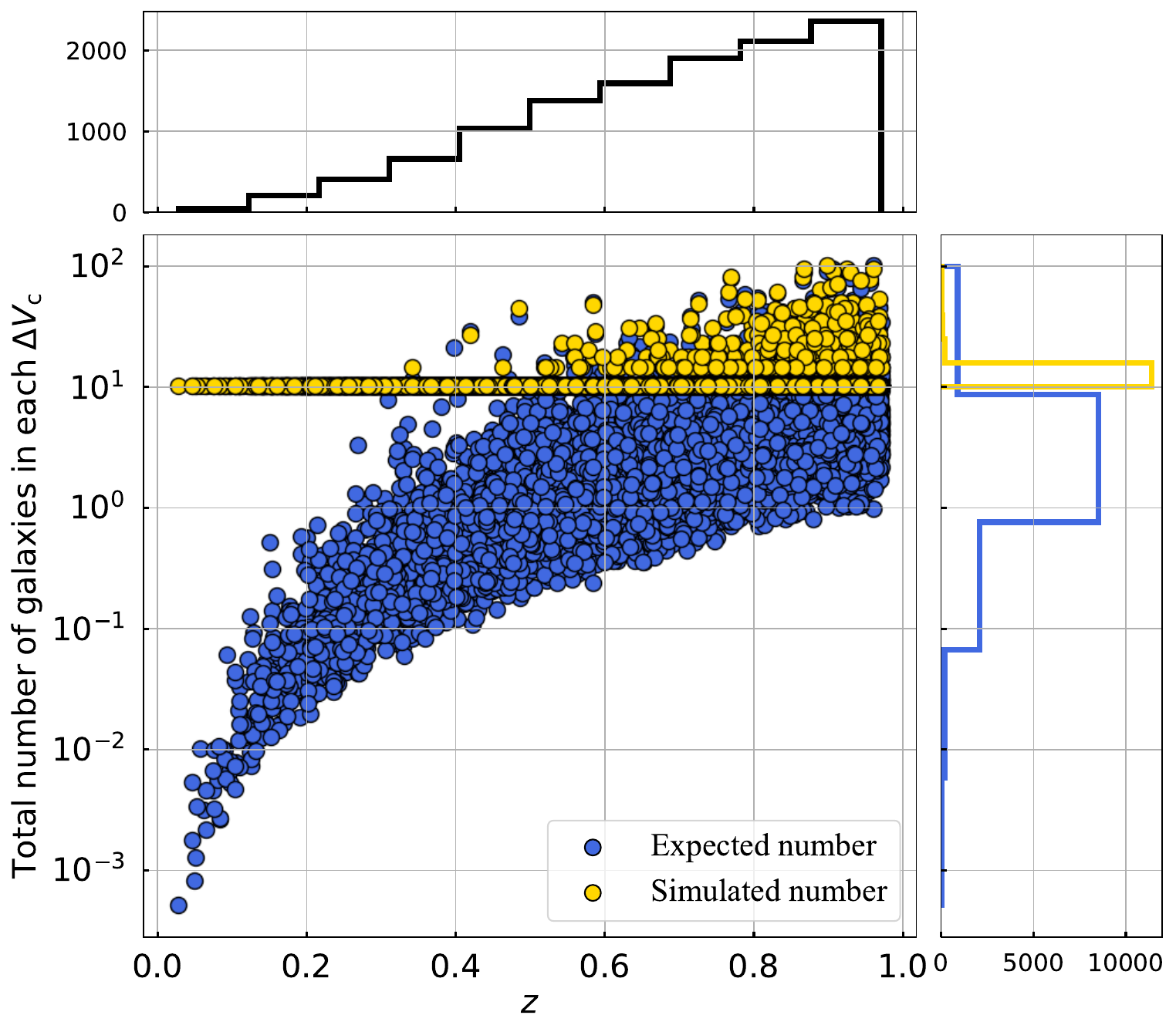}  
  }    
  \caption{The expected number (blue) and the simulated number (yellow) of
  galaxies in each $\Delta V_{\rm c}$ are given in dots, and the detectable merger
  events are summarized with histograms.} 
  \label{fig:number of galaxy}
  \end{figure*}

\subsection{\rm Estimations for a single parameter $H_0$}
\label{sec9}

As mentioned in Sec.~\ref{sec5}, the luminosity distance of the $i$-th simulated
merger event is ${D}_{\rm L}^i$. The corresponding redshifts of possible host
galaxies are $\big\{{z}_{ i,1}, \cdots,{z}_{i,j}, \cdots,{z}_{i,{N}_{\rm gal}^i}
\big\}$. By using Eq.~(\ref{eq1}), we can get each ${H}_0^{ij}$ with ${D}_{\rm
L}^i$ and ${z}_{i,j}$. As noted in Sec.~\ref{sec6}, we have transformed the
redshift uncertainty to the luminosity uncertainty, and then we calculate the
total uncertainty $\sigma_{{H}_0^{ij}}^{\rm tot}= {\lvert {\partial
{H_f(z)}}/{\partial {D_{\rm L}}} \rvert \cdot \lvert 
\sigma_{{D}_{\rm L}^i,j}\rvert}$ of ${H}_0^{ij}$ by using Eq.~(\ref{eq21}) and the chain rule. Note
that $H_f(z)$ is derived by Eq.~(\ref{eq1}). The likelihood of ${H}_0$ constrained from the $i$-th event
 can thus be expressed as
\begin{equation}\label{eq24}
    \begin{aligned}
      {\rm P}(\mathcal{D}_{\rm GW}^i& |{H_0,I})\propto\sum_{j=1}^{{N}_{\rm
      gal}^{i}} \bigg[{M}_{i,j}\times{f}_{\rm pos}(\overline{\theta}_{\rm
      S}^{i,j},\overline{\phi}_{\rm S}^{i,j})\\
      &\times{\frac{1}{\sqrt{2\pi}{\sigma_{{H}_0^{ij}}^{\rm tot}}}{\rm exp}
      \Big(-\frac{({H}_0-{H}_0^{ij})^2}{2({\sigma_{{H}_0^{ij}}^{\rm tot}})^2}
      \Big)} \bigg]\,.
    \end{aligned}
\end{equation}
Finally, we calculate the likelihood of ${H}_{0}$ constrained by all simulated
merger events with Eq.~(\ref{eq10}).  Note that for simplicity, we select the
well-localized simulated merger events with the total number of galaxies in
$\Delta V_{\rm c}$ less than 100, i.e.\ ${N}_{\rm gal}^i \le 100$. As these
events are most constraining, this choice, while speeding up our calculation
enormously, will not change our result in a significant way. 

\begin{figure} 
\centering    
\includegraphics[width=1.0\columnwidth]{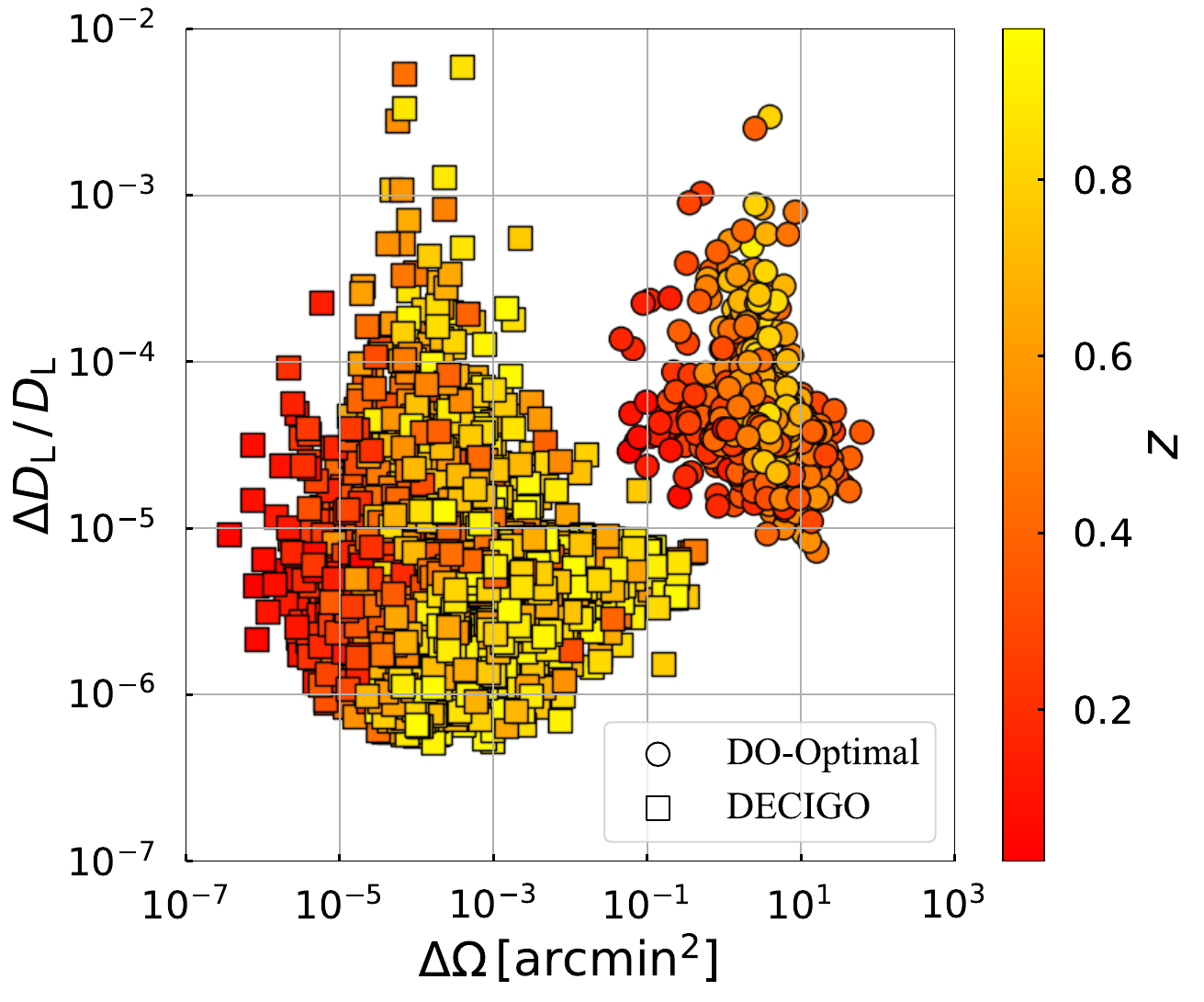}
\caption{The relative uncertainty on the luminosity distance and angular
resolution for DO-Optimal (circles) and DECIGO (squares) in our simulated
samples.}  
\label{fig:resolution}     
\end{figure}

We show in Fig.~\ref{fig:number of galaxy} the relationship between the total
number of simulated galaxies for each merger event and their redshifts. We find
that DO-Optimal can detect well-localized \ac{SBBH} merger events with $z \lesssim
0.5$, while the redshift can be much larger for DECIGO, reaching $z \sim 1$. We
also see that DECIGO has a better performance not only on the total number of
the merger events but also on the ``best-localized'' events. The best-localized
events are defined that there is only one galaxy within $\Delta V_{\rm c}$. As
shown in Fig.~\ref{fig:resolution}, the excellent performance of DECIGO can be
explained by its better distance and angular resolution than DO-Optimal. Fig.~\ref{fig:resolution} 
also shows that, in our samples, DECIGO's minimum values
for the distance relative error and angular resolution can reach $\sim10^{-6}$
and $\sim10^{-6}\, {\rm arcmin}^2$, respectively.

\begin{figure*}
\centering
\subfigure[DO-Optimal]{
    \label{fig:DO-combine-posterior}     
    \includegraphics[width=1\columnwidth]{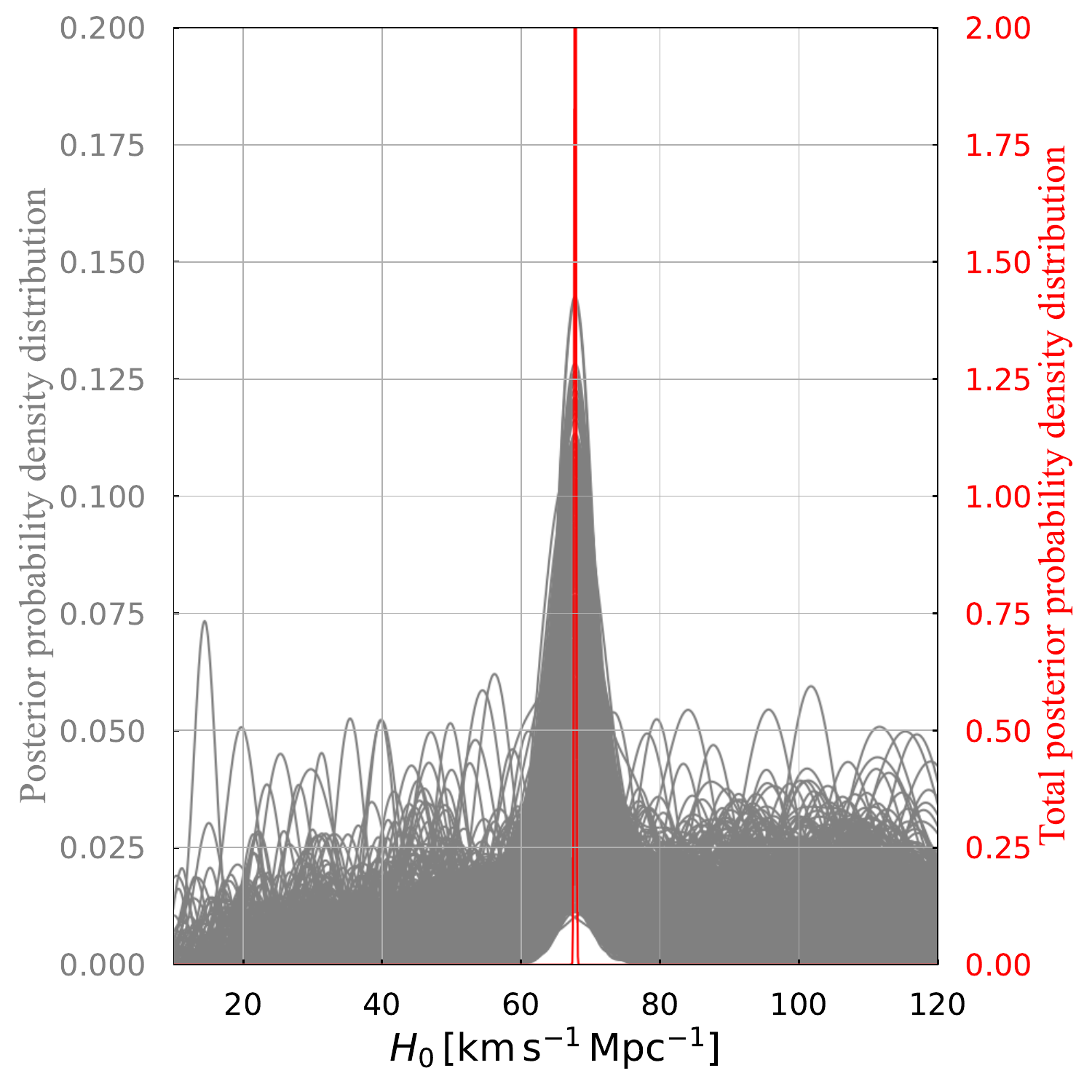}
}     
\subfigure[DECIGO]{ 
    \label{fig:DECIGO-combine-posterior}
    \includegraphics[width=1\columnwidth]{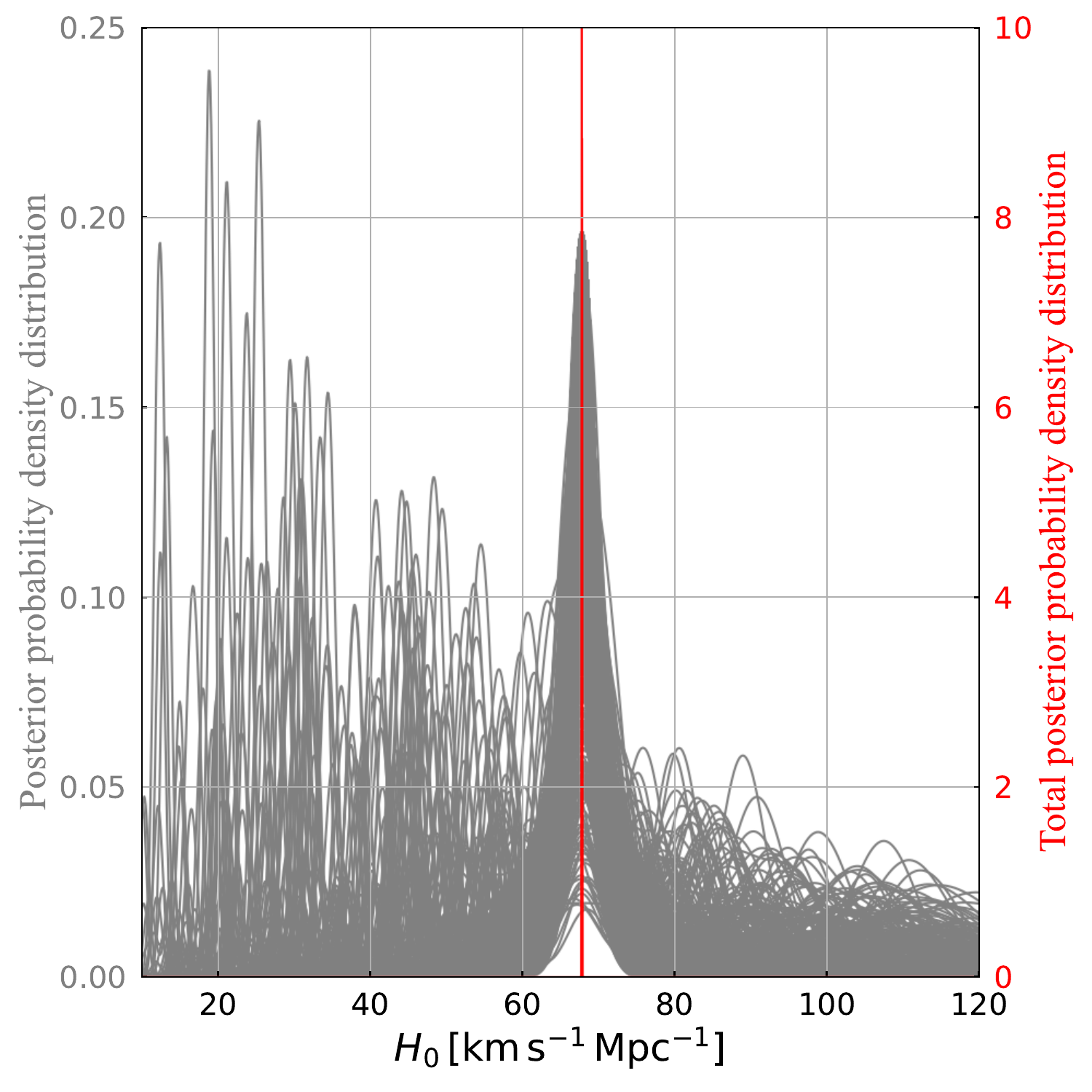}  
}    
\caption{The posterior probability density distributions of ${H}_{0}$ for each
\ac{GW} event (gray) and the total posterior probability density distribution
(red) for DO-Optimal (left) and DECIGO (right).}
\label{fig:combine posterior}
\end{figure*}

\begin{figure*}[htbp]
\centering
\subfigure[DO-Optimal]{
    \label{fig:DO-only-posterior}     
    \includegraphics[width=1\columnwidth]{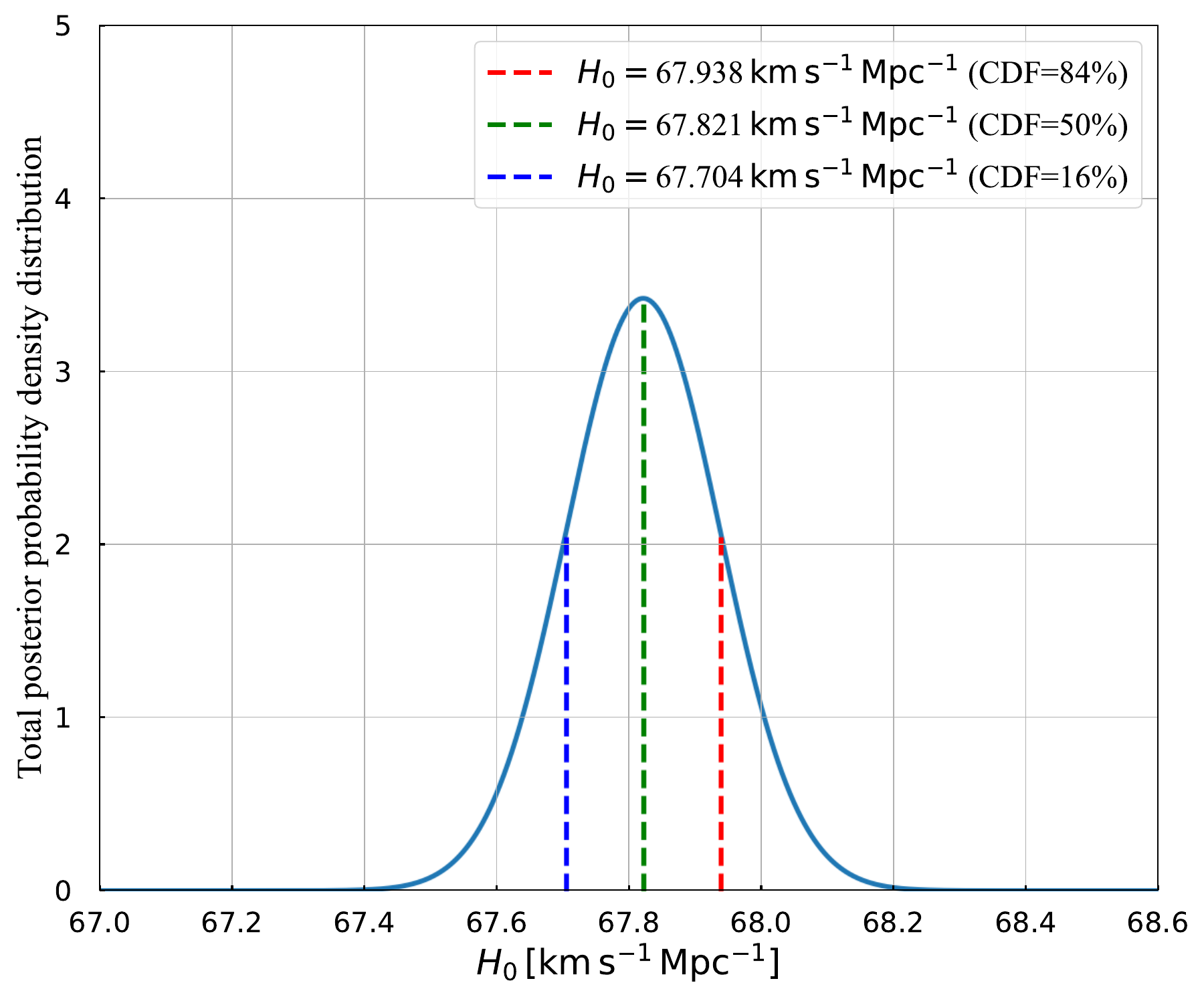}
}
\subfigure[DECIGO]{ 
    \label{fig:DECIGO-only-posterior}
    \includegraphics[width=1\columnwidth]{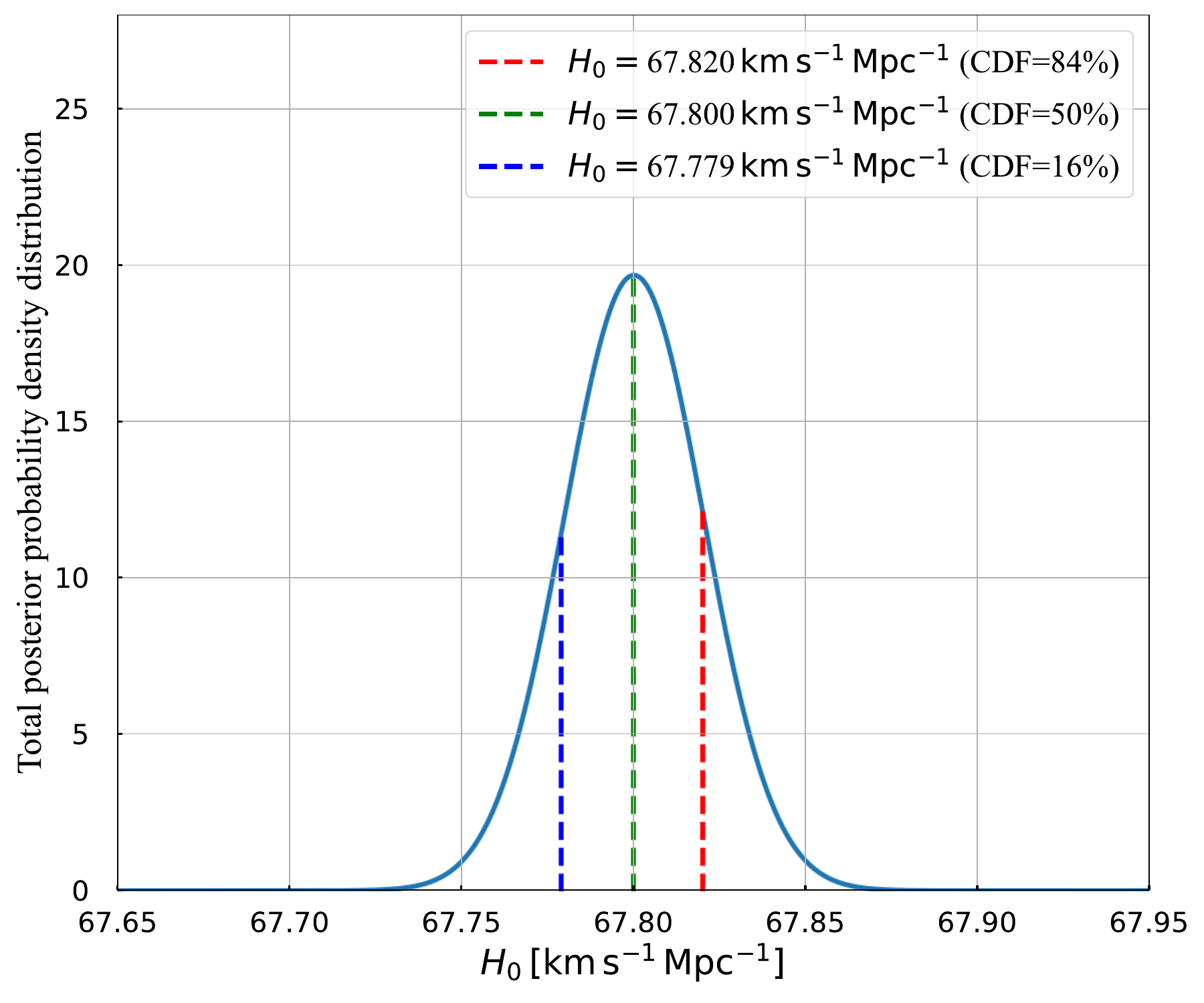}  
}    
\caption{Posterior probability density distributions of ${H}_{0}$ with a 68\%
confidence level from combining all events in DO-Optimal (left) and DECIGO
(right).}
\label{fig:only posterior}
\end{figure*}

\begin{table*}
\begin{center}
\hspace*{-0.05cm}
\begin{tabular}{|cl|cl|clcl|clclcl|}
\hline
\multicolumn{2}{|c|}{\multirow{5}{*}{Detector}}   & \multicolumn{2}{c|}{\multirow{2}{*}{1-parameter}} & \multicolumn{4}{c|}{\multirow{2}{*}{2-parameter}}                                           & \multicolumn{6}{c|}{\multirow{2}{*}{3-parameter}}                                                                                     \\
\multicolumn{2}{|c|}{}                            & \multicolumn{2}{c|}{}                   & \multicolumn{4}{c|}{}                                                             & \multicolumn{6}{c|}{}                                                                                                       \\ \cline{3-14} 
\multicolumn{2}{|c|}{}                            & \multicolumn{2}{c|}{\multirow{3}{*}{$ \frac{\sigma_{H_0}}{H_0}(\%)$}} & \multicolumn{2}{c|}{\multirow{3}{*}{$ \frac{\sigma_{H_0}}{H_0}(\%)$}} & \multicolumn{2}{c|}{\multirow{3}{*}{$\frac{\sigma_{\Omega_{\rm M}}}{\Omega_{\rm M}}(\%)$}} & \multicolumn{2}{c|}{\multirow{3}{*}{$ \frac{\sigma_{H_0}}{H_0}(\%)$}} & \multicolumn{2}{c|}{\multirow{3}{*}{$\frac{\sigma_{\Omega_{\rm M}}}{\Omega_{\rm M}}(\%)$}} & \multicolumn{2}{c|}{\multirow{3}{*}{$\frac{\sigma_{\Omega_\Lambda}}{\Omega_\Lambda}(\%)$}} \\
\multicolumn{2}{|c|}{}                            & \multicolumn{2}{c|}{}                   & \multicolumn{2}{c|}{}                   & \multicolumn{2}{c|}{}                   & \multicolumn{2}{c|}{}                   & \multicolumn{2}{c|}{}                   & \multicolumn{2}{c|}{}                   \\
\multicolumn{2}{|c|}{}                            & \multicolumn{2}{c|}{}                   & \multicolumn{2}{c|}{}                   & \multicolumn{2}{c|}{}                   & \multicolumn{2}{c|}{}                   & \multicolumn{2}{c|}{}                   & \multicolumn{2}{c|}{}                   \\ \hline
\multicolumn{2}{|c|}{\multirow{2}{*}{DO-Optimal}} & \multicolumn{2}{c|}{\multirow{2}{*}{0.17}} & \multicolumn{2}{c|}{\multirow{2}{*}{1.8}} & \multicolumn{2}{c|}{\multirow{2}{*}{44}} & \multicolumn{2}{c|}{\multirow{2}{*}{1.7}} & \multicolumn{2}{c|}{\multirow{2}{*}{86}} & \multicolumn{2}{c|}{\multirow{2}{*}{33}} \\
\multicolumn{2}{|c|}{}                            & \multicolumn{2}{c|}{}                   & \multicolumn{2}{c|}{}                   & \multicolumn{2}{c|}{}                   & \multicolumn{2}{c|}{}                   & \multicolumn{2}{c|}{}                   & \multicolumn{2}{c|}{}                   \\ \hline
\multicolumn{2}{|c|}{\multirow{2}{*}{DECIGO}}     & \multicolumn{2}{c|}{\multirow{2}{*}{0.029}} & \multicolumn{2}{c|}{\multirow{2}{*}{0.14}} & \multicolumn{2}{c|}{\multirow{2}{*}{0.98}} & \multicolumn{2}{c|}{\multirow{2}{*}{0.42}} & \multicolumn{2}{c|}{\multirow{2}{*}{3.3}} & \multicolumn{2}{c|}{\multirow{2}{*}{5.1}} \\
\multicolumn{2}{|c|}{}                            & \multicolumn{2}{c|}{}                   & \multicolumn{2}{c|}{}                   & \multicolumn{2}{c|}{}                   & \multicolumn{2}{c|}{}                   & \multicolumn{2}{c|}{}                   & \multicolumn{2}{c|}{}                   \\ \hline
\multicolumn{14}{|c|}{\multirow{2}{*}{Using a realistic simulated galaxy catalog from the TAO}}\\ 
\multicolumn{14}{|c|}{}\\\hline
\multicolumn{2}{|c|}{\multirow{2}{*}{DECIGO}}     & \multicolumn{2}{c|}{\multirow{2}{*}{0.032}} & \multicolumn{2}{c|}{\multirow{2}{*}{0.18}} & \multicolumn{2}{c|}{\multirow{2}{*}{1.3}} & \multicolumn{2}{c|}{\multirow{2}{*}{0.47}} & \multicolumn{2}{c|}{\multirow{2}{*}{4.6}} & \multicolumn{2}{c|}{\multirow{2}{*}{6.6}} \\
\multicolumn{2}{|c|}{}                            & \multicolumn{2}{c|}{}                   & \multicolumn{2}{c|}{}                   & \multicolumn{2}{c|}{}                   & \multicolumn{2}{c|}{}                   & \multicolumn{2}{c|}{}                   & \multicolumn{2}{c|}{}                   \\ \hline
\end{tabular}
\caption{Constraints on the cosmological parameters with DO-Optimal and DECIGO.}
\label{tab:Constraints for cosmological parameter by DO-Optimal and DECIGO}
\end{center}
\end{table*}

\begin{table*}
\begin{center}
\begin{tabular}{|clclclclcllcllclllllclllllclllll|}
\hline
\multicolumn{2}{|c|}{\multirow{6}{*}{Detector}}   & \multicolumn{6}{c|}{\multirow{2}{*}{$N_{\rm tot}$}}                                                                                   & \multicolumn{6}{c|}{\multirow{4}{*}{\begin{tabular}[c]{@{}c@{}}${N}_{\rm gal} \le 100$\\ (well-localized)\end{tabular}}} & \multicolumn{18}{c|}{\multirow{2}{*}{SNR>10}}                                                                                                                                           \\
\multicolumn{2}{|c|}{}                            & \multicolumn{6}{c|}{}                                                                                                    & \multicolumn{6}{c|}{}                                                                                          & \multicolumn{18}{c|}{}                                                                                                                                                                               \\ \cline{3-8} \cline{15-32} 
\multicolumn{2}{|c|}{}                            & \multicolumn{6}{c|}{\multirow{2}{*}{11696}}                                                                                   & \multicolumn{6}{c|}{}                                                                                          & \multicolumn{6}{c|}{\multirow{4}{*}{{${N}_{\rm gal} \le 10$}}}    & \multicolumn{6}{c|}{\multirow{4}{*}{{${N}_{\rm gal} \le 3$}}}    & \multicolumn{6}{c|}{\multirow{4}{*}{\begin{tabular}[c]{@{}c@{}}${N}_{\rm gal} = 1$\\ (best-localized)\end{tabular}}} \\
\multicolumn{2}{|c|}{}                            & \multicolumn{6}{c|}{}                                                                                                    & \multicolumn{6}{c|}{}                                                                                          & \multicolumn{6}{c|}{}                     & \multicolumn{6}{c|}{}                     & \multicolumn{6}{c|}{}                                                                                        \\ \cline{3-14}
\multicolumn{2}{|c|}{}                            & \multicolumn{2}{c|}{\multirow{2}{*}{SNR>10}} & \multicolumn{2}{c|}{\multirow{2}{*}{SNR>50}} & \multicolumn{2}{c|}{\multirow{2}{*}{SNR>100}} & \multicolumn{3}{c|}{\multirow{2}{*}{SNR>10}}                 & \multicolumn{3}{c|}{\multirow{2}{*}{SNR>50}}                & \multicolumn{6}{c|}{}                     & \multicolumn{6}{c|}{}                     & \multicolumn{6}{c|}{}                                                                                        \\
\multicolumn{2}{|c|}{}                            & \multicolumn{2}{c|}{}                  & \multicolumn{2}{c|}{}                  & \multicolumn{2}{c|}{}                  & \multicolumn{3}{c|}{}                                  & \multicolumn{3}{c|}{}                                 & \multicolumn{6}{c|}{}                     & \multicolumn{6}{c|}{}                     & \multicolumn{6}{c|}{}                                                                                        \\ \hline
\multicolumn{2}{|c|}{\multirow{2}{*}{DO-Optimal}} & \multicolumn{2}{c|}{\multirow{2}{*}{11694}} & \multicolumn{2}{c|}{\multirow{2}{*}{8210}} & \multicolumn{2}{c|}{\multirow{2}{*}{3741}} & \multicolumn{3}{c|}{\multirow{2}{*}{860}}                 & \multicolumn{3}{c|}{\multirow{2}{*}{719}}                & \multicolumn{6}{c|}{\multirow{2}{*}{329}}    & \multicolumn{6}{c|}{\multirow{2}{*}{194}}    & \multicolumn{6}{c|}{\multirow{2}{*}{133}}                                                                       \\
\multicolumn{2}{|c|}{}                            & \multicolumn{2}{c|}{}                  & \multicolumn{2}{c|}{}                  & \multicolumn{2}{c|}{}                  & \multicolumn{3}{c|}{}                                  & \multicolumn{3}{c|}{}                                 & \multicolumn{6}{c|}{}                     & \multicolumn{6}{c|}{}                     & \multicolumn{6}{c|}{}                                                                                        \\ \hline
\multicolumn{2}{|c|}{\multirow{2}{*}{DECIGO}}     & \multicolumn{2}{c|}{\multirow{2}{*}{11696}} & \multicolumn{2}{c|}{\multirow{2}{*}{11696}} & \multicolumn{2}{c|}{\multirow{2}{*}{11693}} & \multicolumn{3}{c|}{\multirow{2}{*}{11689}}                 & \multicolumn{3}{c|}{\multirow{2}{*}{11689}}                & \multicolumn{6}{c|}{\multirow{2}{*}{11626}}    & \multicolumn{6}{c|}{\multirow{2}{*}{11491}}    & \multicolumn{6}{c|}{\multirow{2}{*}{11120}}                                                                       \\
\multicolumn{2}{|c|}{}                            & \multicolumn{2}{c|}{}                  & \multicolumn{2}{c|}{}                  & \multicolumn{2}{c|}{}                  & \multicolumn{3}{c|}{}                                  & \multicolumn{3}{c|}{}                                 & \multicolumn{6}{c|}{}                     & \multicolumn{6}{c|}{}                     & \multicolumn{6}{c|}{}                                                                                        \\ \hline
\multicolumn{32}{|c|}{\multirow{2}{*}{Using a realistic simulated galaxy catalog from the TAO}}                                                                                                                                                                                                                                                                                                                                                                                                                                                             \\
\multicolumn{32}{|c|}{}                                                                                                                                                                                                                                                                                                                                                                                                                                                                              \\ \hline
\multicolumn{2}{|c|}{\multirow{8}{*}{DECIGO}}     & \multicolumn{6}{c|}{\multirow{2}{*}{$N_{\rm sel}$}}                                                                                      & \multicolumn{3}{c|}{\multirow{8}{*}{11293}}               & \multicolumn{3}{c|}{\multirow{8}{*}{11293}}               & \multicolumn{6}{c|}{\multirow{8}{*}{10396}} & \multicolumn{6}{c|}{\multirow{8}{*}{8584}} & \multicolumn{6}{c|}{\multirow{8}{*}{5611}}                                                                    \\
\multicolumn{2}{|c|}{}                            & \multicolumn{6}{c|}{}                                                                                                       & \multicolumn{3}{c|}{}                                  & \multicolumn{3}{c|}{}                                 & \multicolumn{6}{c|}{}                     & \multicolumn{6}{c|}{}                     & \multicolumn{6}{c|}{}                                                                                        \\ \cline{3-8}
\multicolumn{2}{|c|}{}                            & \multicolumn{6}{c|}{\multirow{2}{*}{11396}}                                                                                      & \multicolumn{3}{c|}{}                                  & \multicolumn{3}{c|}{}                                 & \multicolumn{6}{c|}{}                     & \multicolumn{6}{c|}{}                     & \multicolumn{6}{c|}{}                                                                                        \\
\multicolumn{2}{|c|}{}                            & \multicolumn{6}{c|}{}                                                                                                       & \multicolumn{3}{c|}{}                                  & \multicolumn{3}{c|}{}                                 & \multicolumn{6}{c|}{}                     & \multicolumn{6}{c|}{}                     & \multicolumn{6}{c|}{}                                                                                        \\ \cline{3-8}
\multicolumn{2}{|c|}{}                            & \multicolumn{2}{c|}{\multirow{2}{*}{SNR>10}} & \multicolumn{2}{c|}{\multirow{2}{*}{SNR>50}} & \multicolumn{2}{c|}{\multirow{2}{*}{SNR>100}} & \multicolumn{3}{c|}{}                                  & \multicolumn{3}{c|}{}                                 & \multicolumn{6}{c|}{}                     & \multicolumn{6}{c|}{}                     & \multicolumn{6}{c|}{}                                                                                        \\
\multicolumn{2}{|c|}{}                            & \multicolumn{2}{c|}{}                   & \multicolumn{2}{c|}{}                   & \multicolumn{2}{c|}{}                   & \multicolumn{3}{c|}{}                                  & \multicolumn{3}{c|}{}                                 & \multicolumn{6}{c|}{}                     & \multicolumn{6}{c|}{}                     & \multicolumn{6}{c|}{}                                                                                        \\ \cline{3-8}
\multicolumn{2}{|c|}{}                            & \multicolumn{2}{c|}{\multirow{2}{*}{11396}}  & \multicolumn{2}{c|}{\multirow{2}{*}{11396}}  & \multicolumn{2}{c|}{\multirow{2}{*}{11394}}  & \multicolumn{3}{c|}{}                                  & \multicolumn{3}{c|}{}                                 & \multicolumn{6}{c|}{}                     & \multicolumn{6}{c|}{}                     & \multicolumn{6}{c|}{}                                                                                        \\
\multicolumn{2}{|c|}{}                            & \multicolumn{2}{c|}{}                   & \multicolumn{2}{c|}{}                   & \multicolumn{2}{c|}{}                   & \multicolumn{3}{c|}{}                                  & \multicolumn{3}{c|}{}                                 & \multicolumn{6}{c|}{}                     & \multicolumn{6}{c|}{}                     & \multicolumn{6}{c|}{}                                                                                        \\ \hline
\end{tabular}
\caption{The number of \ac{GW} events with different SNR thresholds and ${N}_{\rm gal}$.}
\label{tab:Numberevent}
\end{center}
\end{table*}

We can see in Fig.~\ref{fig:combine posterior} that there are multiple peaks on
the posterior probability density distributions for each merger event, given by
gray lines. For each gray line, among the peaks it has, there is always 
a peak near the true value, $H_{0}=67.8\,{\rm km\,s^{-1}\,Mpc^{-1}}$ in 
our injection.
In addition, the well-localized \ac{GW} events $({N}_{\rm gal}^i \le 100)$ tend to
have better constraints on $H_{0}$, and have made a major contribution. For more
comparisons between DO-Optimal and DECIGO, we show the $1$-$\sigma$ range of the
combined posterior probability density distributions in Fig.~\ref{fig:only
posterior}. The given values of \ac{CDF} are the medians, along with its $16\%$ and $84\%$ quantiles. We can see that the relative uncertainties of $H_{0}$ for DO-Optimal
and DECIGO are about $0.17\%$ and $0.029\%$, respectively. Therefore, both
DO-Optimal and DECIGO can constrain $H_0$ quite well. Between them, DECIGO
performs even better. It is not only related to DECIGO's excellent positioning
capability but also its larger number of detectable merger events (cf.
Figs.~\ref{fig:number of galaxy}--\ref{fig:resolution} and Table \ref{tab:Numberevent}).

\subsection{\rm Estimations for multiple parameters}
\label{sec10}

For simplicity, we uniformly select \ac{GW} events with ${N}_{\rm gal}^i \le 3$.
We have verified with ${N}_{\rm gal}^i \le 10$ that the results are consistent.
Thus we only illustrate our results with ${N}_{\rm gal}^i \le 3$ in this
subsection.

\begin{figure*}[htbp]
\centering
\subfigure[DO-Optimal]{
    \label{fig:DO-2parameter}     
    \includegraphics[width=0.9\columnwidth]{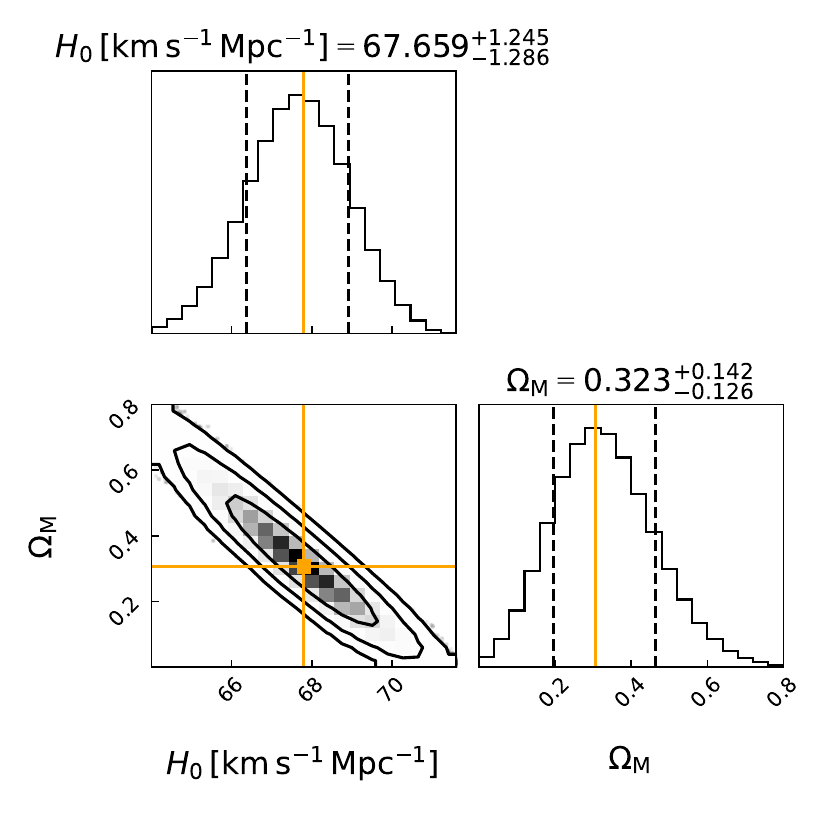}
}     
\subfigure[DECIGO]{ 
    \label{fig:DECIGO-2parameter}
    \includegraphics[width=0.9\columnwidth]{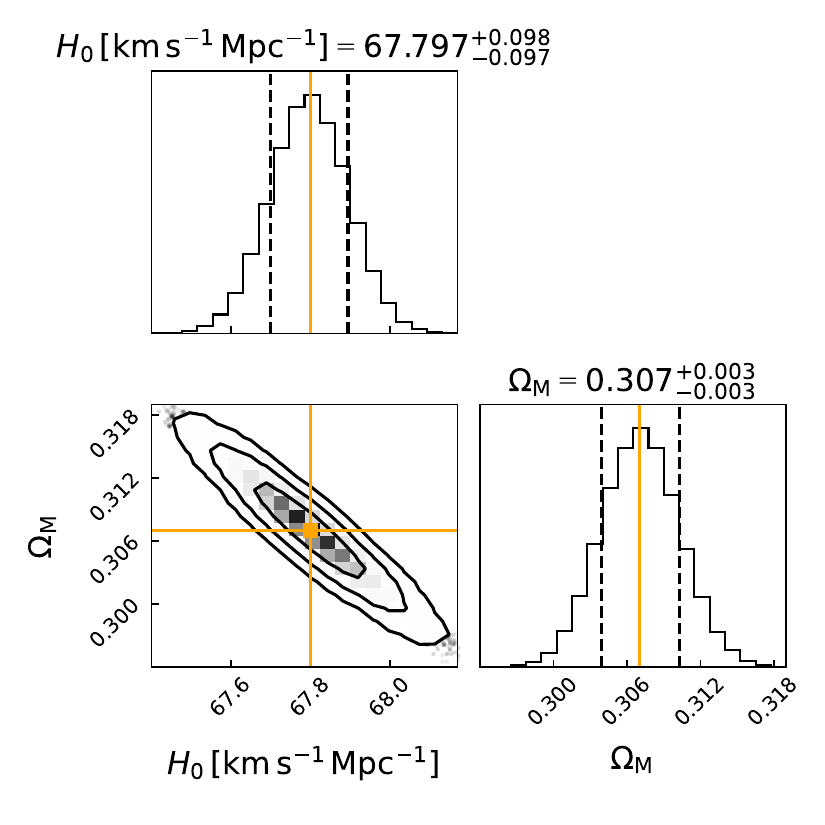}  
}    
\caption{Two-parameter estimations for $\big\{H_0,\Omega_{\rm M}\big\}$ with
DO-Optimal (left) and DECIGO (right).}
\label{fig:2parameter}
\end{figure*}

\begin{figure*}[htbp]
\centering
\subfigure[DO-Optimal]{
    \label{fig:DO-3parameter}     
    \includegraphics[width=1\columnwidth]{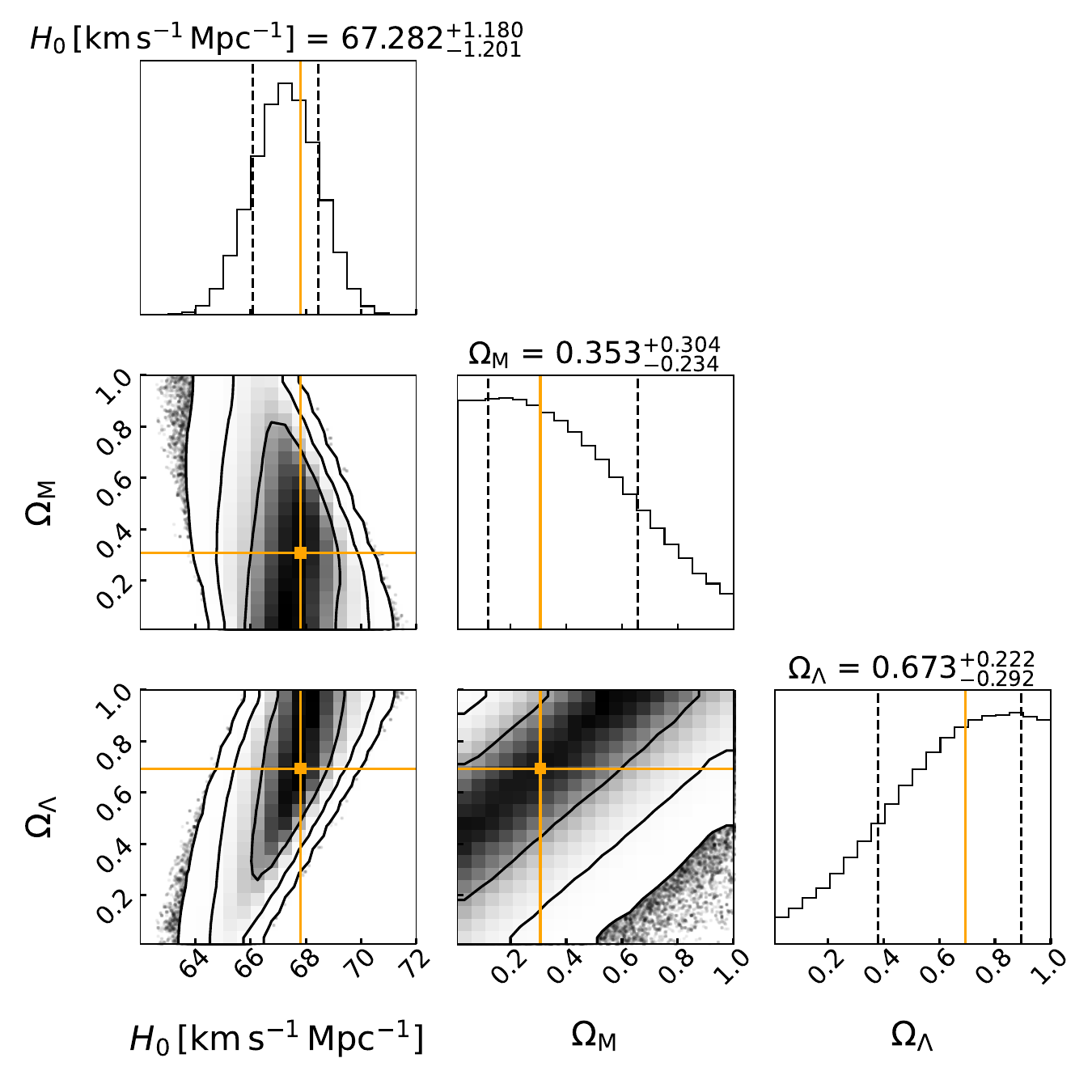}
}     
\subfigure[DECIGO]{ 
    \label{fig:DECIGO-3parameter}
    \includegraphics[width=1\columnwidth]{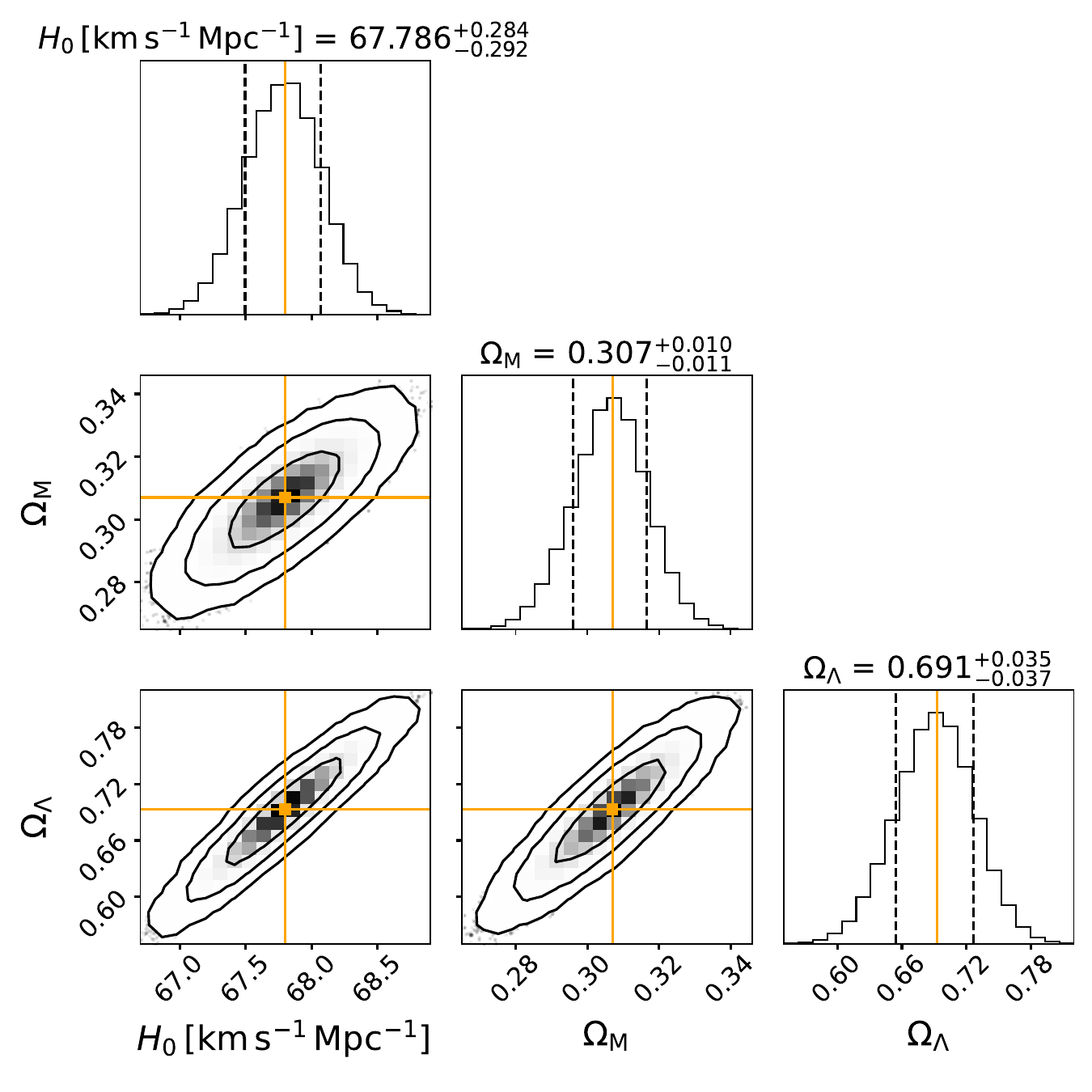}  
}    
\caption{Three-parameter estimations for $\big\{H_0,{\rm
\Omega_M,\Omega_\Lambda}\big\}$ with DO-Optimal (left) and DECIGO (right).}
\label{fig:3parameter}
\end{figure*}

We first consider two parameters $\big\{H_0,\Omega_{\rm M}\big\}$ with
$\Omega_\Lambda=1-\Omega_{\rm M}$ in Eq.~(\ref{eq2}). We find in
Fig.~\ref{fig:2parameter} that DECIGO has better constraints on
$\big\{H_0,\Omega_{\rm M}\big\}$ with relative errors of $\sim0.14\%$ and
$\sim0.98\%$, respectively. In contrast, DO-Optimal can constrain $H_0$ with a
relative error of $\sim1.8\%$, while it performs worse on $\Omega_{\rm M}$ with
a relative error of $\sim44\%$.

In addition to the results above, we further consider treating $\Omega_\Lambda$
as an independent variable and then performing parameter estimations on
$\mathcal{H}$ with three parameters. Fig.~\ref{fig:DO-3parameter} shows the
estimation results for the two detectors. We see in Fig.~\ref{fig:DO-3parameter}
that DO-Optimal can constrain $H_0$ well with a relative error of $\sim1.7\%$,
but has basically no constraints on $\Omega_{\rm M}$ and $\Omega_\Lambda$ with
relative errors reaching $\sim86\%$ and $\sim33\%$, respectively. This can be
explained well by the low localization accuracy of the high-redshift events for
DO-Optimal. In contrast, DECIGO still has a much better performance on the
parameter estimations, as shown in Fig.~\ref{fig:DECIGO-3parameter}. The
relative errors of $\big\{H_0,{\rm \Omega_M,\Omega_\Lambda}\big\}$ can reach
$\sim0.42\%$, $\sim3.3\%$, and $\sim5.1\%$, respectively.

Comparing the 2-parameter and 3-parameter constraints from the two detectors, we
conclude that DECIGO behaves much better than DO-Optimal. Both detectors can
constrain $H_0$ well, while for $\Omega_{\rm M}$ and $\Omega_{\Lambda}$, only
DECIGO can reach a significant result with  relative errors $\le6\%$. The
differences between DECIGO and DO-Optimal are mainly due to the different
localization capabilities, especially for the high-redshift events. DECIGO can
detect well-localized merger events even at $z=1$, significantly improving the
constraints. Moreover, $\Omega_{\rm M}$ and $\Omega_{\Lambda}$ dominate in the
high-redshift regime based on Eq.~(\ref{eq1}), which determines that it is hard
for DO-Optimal to give more precise measurements of $\Omega_{\rm M}$ and
$\Omega_{\Lambda}$.

It is worth noting that for the same detector, the single parameter constraints
can be much better than the results of the multiple parameters.  The increase in
relative error can be explained by the degrees of freedom. We summarize our
marginalized constraints in Table \ref{tab:Constraints for cosmological
parameter by DO-Optimal and DECIGO} and the number of \ac{GW} events 
under different cutoffs in calculations in Table \ref{tab:Numberevent}.

\subsection{\rm Estimations with a realistic simulated galaxy catalog}
\label{secadd}

We adopt the \ac{MDPL} cosmological simulation \citep[][]{10.1093/mnras/stw248} 
and the \ac{SAGE} model \citep[][]{Croton:2016etl} 
from the \ac{TAO} \footnote{https://tao.asvo.org.au/tao/.} to 
generate a realistic simulated galaxy catalog. The \ac{MDPL} simulation assumes a Planck cosmology 
\citep{Planck:2015fie}, and the \ac{TAO} is a publicly available codebase that runs on the 
dark matter halo trees of a cosmological N-body simulation. In addition to the locations of the galaxies,
the \ac{MDPL} also provide the luminosity information of the galaxies according to \citet{Croton:2016etl,Conroy:2008dx}.

All the information obtained from the simulated catalog includes luminosity distance, redshift, 
mass, location, and apparent magnitude in the K-band. Then we adopt a mass-weighted random selection 
of $N_{\rm tot}$ host galaxies. To be more realistic, we also simulate 
the \ac{EM} selection effect. Following \citet{Zhu:2021aat}, the probability that a galaxy 
with luminosity $\mathcal{L}$ (in the unit of sun's luminosity) can be observed is defined as
\begin{equation}\label{eq25}
  \begin{aligned}
    {\mathcal{P}({\rm lg}\mathcal{L})=\int_{{\rm lg}\mathcal{L}_{\rm 
    limit}}^{\infty} \mathcal{N}\big[\mathcal{L},(\sigma_{{\rm lg}\mathcal{L}})^2\big] \,d\mathcal{L}' }\,,
  \end{aligned}
\end{equation}
where ${\rm lg}\mathcal{L}_{\rm limit}=\frac{{\rm m}_\odot-{\rm m_{\rm limit}}-5}{2.5}+2
{\rm lg}\big(\frac{D_{\rm L}}{1 {\rm pc}}\big)$ with the absolute magnitude of the 
sun ${\rm m}_\odot=+4.8\,{\rm mag}$ and the limiting apparent magnitude 
${\rm m_{\rm limit}}=+24\,{\rm mag}$. We adopt the measurement error of limiting 
luminosity $\sigma_{{\rm lg}\mathcal{L}}=0.04$ \citep{Zhu:2021aat,
SDSS:2000hjo,EUCLID:2011zbd}. As shown in Table \ref{tab:Numberevent}, the total number of \ac{SBBH} merger 
events is reduced to $N_{\rm sel}=11396$ due to the \ac{EM} selection effect. 
Considering that the calculation is very time-consuming, we only use \ac{DECIGO} as a 
representative to constrain the cosmological parameters and compare them with the 
previous results in Sec.~\ref{sec9} and \ref{sec10}.

\begin{figure}[htbp]
  \subfigure[Estimation for $\big\{H_0\big\}$]{
      \label{fig:newDECIGO-only-posterior}     
      \includegraphics[width=0.9\columnwidth]{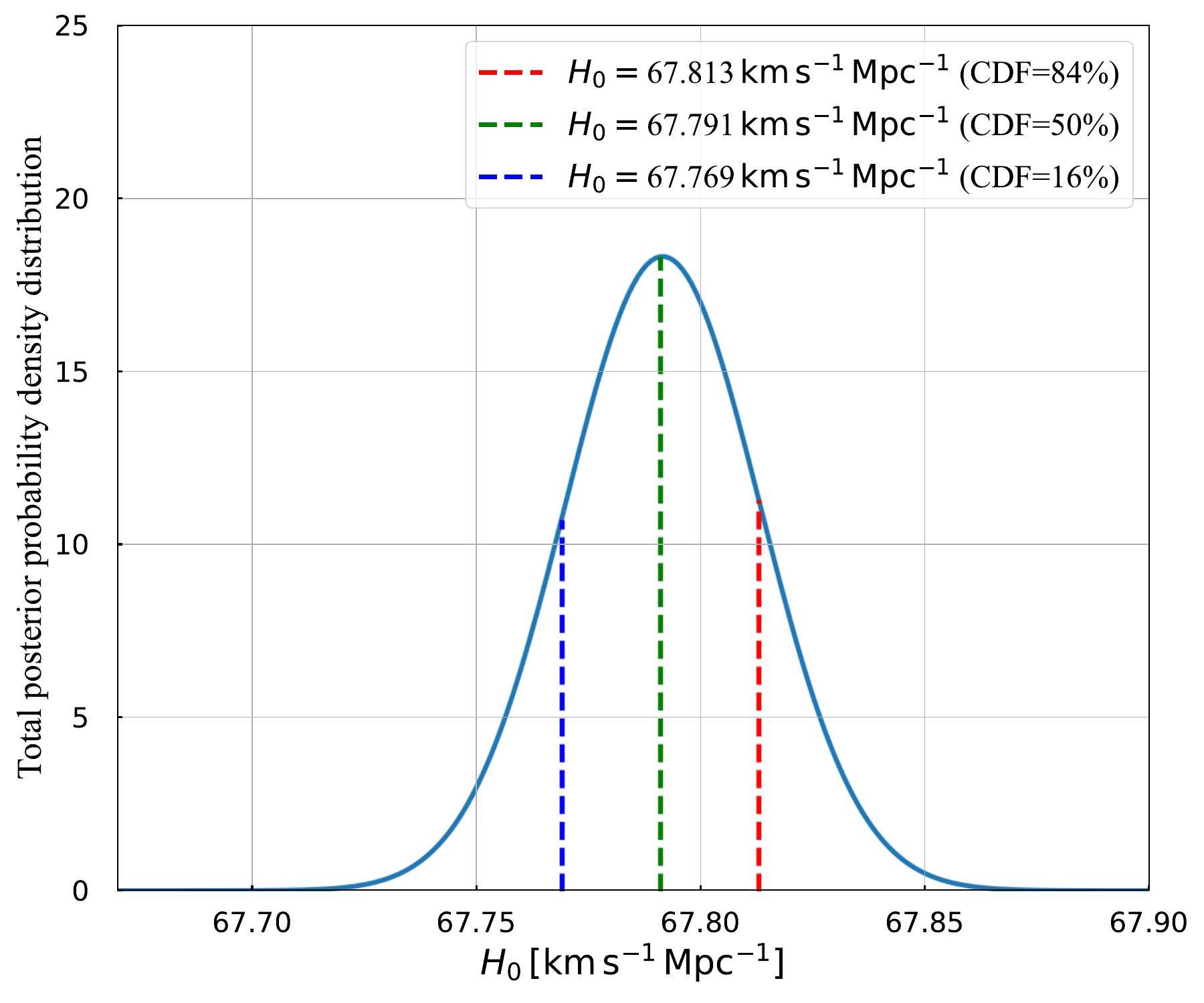}
  }     
  \subfigure[Estimations for $\big\{H_0,\Omega_{\rm M}\big\}$]{ 
      \label{fig:newDECIGO-2parameter}
      \includegraphics[width=0.9\columnwidth]{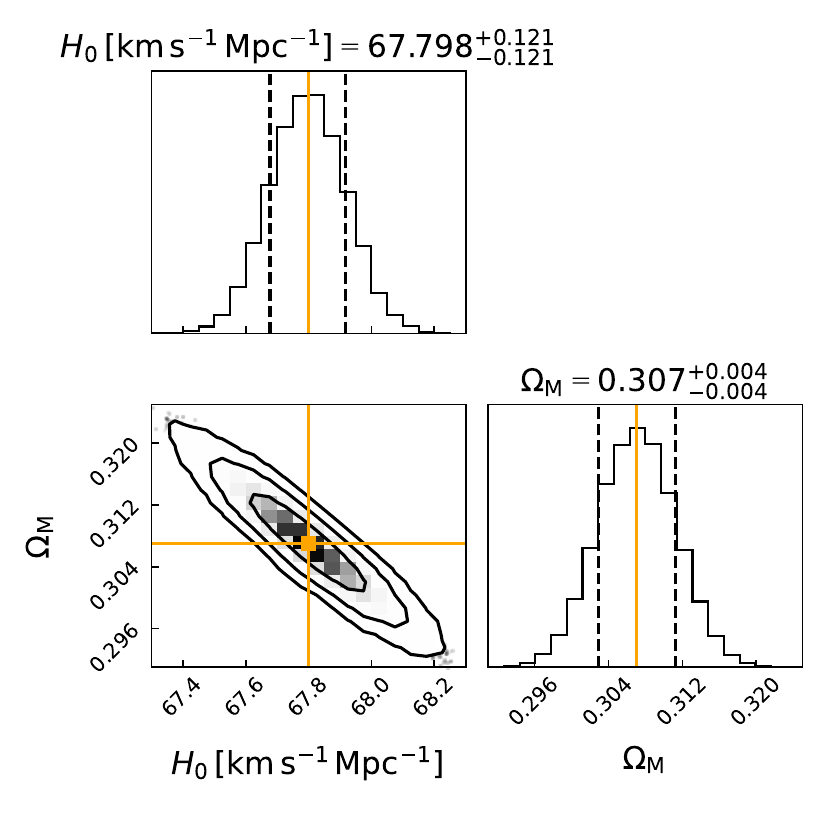}  
  } 
  \subfigure[Estimations for $\big\{H_0,\Omega_{\rm M},\Omega_\Lambda\big\}$]{ 
      \label{fig:newDECIGO-3parameter}
      \includegraphics[width=0.9\columnwidth]{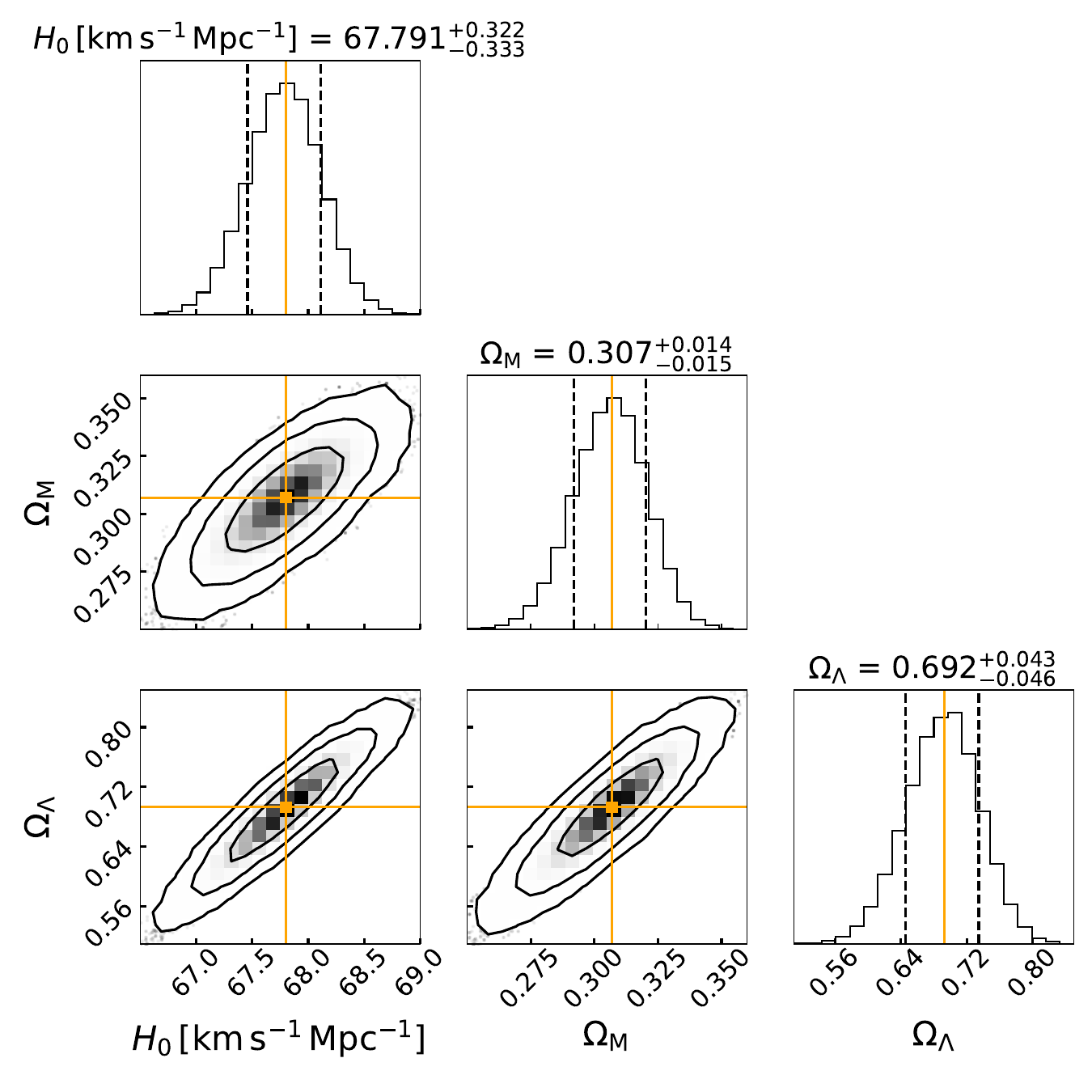}  
  }       
  \caption{Parameter estimations with DECIGO by using a realistic 
  simulated galaxy catalog from the \ac{TAO}.}
  \label{fig:newdecigo}
\end{figure}

We adopt the same method as in Sec.~\ref{sec5} to constrain the 
cosmological parameters. Note that when we calculate the position weights of 
galaxies in each $\Delta V_{\rm c}$, to be more realistic, we have added a random bias to the location of the error volume's 
center relative to the host galaxy. The results are shown in 
Fig.~\ref{fig:newdecigo} and summarized in Table \ref{tab:Constraints for cosmological
parameter by DO-Optimal and DECIGO}. We find that even considering the \ac{EM} 
selection effect with a realistic simulated galaxy catalog, DECIGO can also 
constrain the cosmological parameters well. The relative uncertainties of $H_0$ 
for \ac{DECIGO} is $\sim0.032\%$; $\big\{H_0,{\rm \Omega_M}\big\}$ is about 
$\sim0.18\%$ and $\sim1.3\%$. For the 3-parameter $\big\{H_0,{\rm \Omega_M,\Omega_\Lambda}
\big\}$ estimations, \ac{DECIGO} can reach $\sim0.47\%$, $\sim4.6\%$, 
and $\sim6.6\%$, respectively. In addition, we find that DECIGO's constraints 
on cosmological parameters are consistent (a little worse) with the previous results in 
Sec.~\ref{sec9} and \ref{sec10}, which indicates that the \ac{EM} selection effect has 
little influence on the results of DECIGO's constraints. Besides, 11396 \ac{GW} 
events with SNR>10 also provide a lot of best-localized events 
(5611 \ac{GW} events).

\section{Conclusion and discussion}
\label{sec11}

Up to now, most of the detected \ac{GW} events belong to the dark sirens in the
context of cosmology. In the future, space-borne decihertz \ac{GW} detectors can
provide  excellent localization capability for numerous detectable merger
events. Therefore, it is possible to use these events to constrain cosmological
parameters precisely by a statistic method. In this work, we explore how well
the space-borne decihertz \ac{GW} detectors (DO-Optimal and DECIGO) can
constrain the cosmological parameters with simulated dark-siren events and
galaxy catalogs. We have considered various redshift and luminosity distance
errors and performed parameter estimations with single/multiple parameters for
DO-Optimal and \ac{DECIGO}. All the results are listed in Table
\ref{tab:Constraints for cosmological parameter by DO-Optimal and DECIGO}, and
we find that both DECIGO and DO-Optimal can constrain $H_0$ well. For
2-parameter/3-parameter estimations, DECIGO can still constrain $\Omega_{\rm M}$
and $\Omega_{\Lambda}$ well, while DO-Optimal cannot. The main reason is in
DECIGO's better localization accuracy compared with  DO-Optimal. More recently, 
\citet{Seymour:2022teq} has shown that a space-borne decihertz detector can enhance the sensitivity 
of a ground network by about a factor of 3 for the cosmological parameter estimations with dark sirens. 
The precise measurements of $H_0$ with space-borne decihertz \ac{GW} detectors 
can shed light on the $H_0$ tension. Early measurements constrained the relative 
error of $H_0$ to be larger than a few percents, while space-borne decihertz 
\ac{GW} detectors can reach a better level. With the help of these detectors, 
we can use \ac{GW} as another means to measure $H_0$ precisely and help us 
judge whether the two sets of results in the previous work for $H_0$ are valid.

Note that for realistic cosmological parameter estimations, galaxy 
catalogs would be incomplete due to the \ac{EM} selection effects like the 
Malmquist bias \cite{1922MeLuF.100....1M}. Thus, in our work, we discuss 
the effect of Malmquist bias on the constraints of the cosmological 
parameters using \ac{DECIGO}. We find that 
even taking into account the Malmquist bias, there are also many best-localized 
\ac{GW} events and high-accuracy constraints on cosmological parameters. 
Moreover, several works have derived a correction term into the Bayesian framework 
to eliminate such a selection bias
\citep[][]{LIGOScientific:2017adf,Zhu:2021aat,Chen:2017rfc,Mandel:2018mve}. 
Therefore, our results in this paper can still be used to provide meaningful references 
and helpful inputs for upcoming space-borne decihertz \ac{GW} projects.
In addition, the incompleteness of the galaxy catalog can be compensated by the
\ac{GW}-triggered deeper field galaxy surveys
\citep[][]{Bartos:2014spa,Chen:2015nlv,Klingler:2019fbl}. Moreover, one of the
solutions to the lack of redshift information is to increase the \ac{EM}
follow-up observation. Several works have been devoted to the best searching
strategy with ground-based \ac{GW} observatories
\citep[][]{Gehrels:2015uga,Rosswog:2016dhy,Cowperthwaite:2018gmx,Liu:2020bgc}
and the early warnings of \ac{EM} follow-up observations from decihertz \ac{GW}
observatories \citep{Kang:2022nmz, Liu:2022mcd}. We will investigate these
effects in the future.  Finally, we have only considered circular orbits for
\ac{SBBH}s. Since the eccentricity can provide more astrophysical information
\citep[][]{Chen:2017gfm,Liu:2019jpg,Gerosa:2019dbe,Zhang:2019puc,Nishizawa:2016jji},
it is worth investigating whether the \ac{SBBH} system with eccentricity will
impact the estimation of the cosmological parameters \citep{Zhu:2021bpp}.

\section*{Acknowledgements}

We thank Liang-Gui Zhu and Hui Tong for helpful discussions.  This work was
supported by the National Natural Science Foundation of China (11991053,
12173104, 11975027, 11721303), the Guangdong Major Project of Basic and Applied
Basic Research (2019B030302001), the National SKA Program of China
(2020SKA0120300), the Max Planck Partner Group Program funded by the Max Planck
Society, and the High-Performance Computing Platform of Peking University.  Y.K.
acknowledges the Hui-Chun Chin and Tsung-Dao Lee Chinese Undergraduate Research
Endowment (Chun-Tsung Endowment) at Peking University. 



\bibliographystyle{apsrev4-1}
\bibliography{reference}





\end{document}